\documentclass{article}

\usepackage{arxiv}

\usepackage[utf8]{inputenc}
\usepackage[T1]{fontenc}
\usepackage[colorlinks=true,linkcolor=blue,citecolor=blue]{hyperref}
\usepackage[labelfont=bf]{caption}
\usepackage{subcaption}
\usepackage{url} 
\usepackage{booktabs}
\usepackage{amsmath}
\usepackage{amsfonts}
\usepackage{amssymb}
\usepackage{nicefrac}
\usepackage{microtype}
\usepackage{lipsum}
\usepackage{graphicx}
\usepackage{float}
\usepackage{array}
\usepackage{bm}
\usepackage{mathrsfs,xfrac}
\usepackage{mathtools}
\usepackage{xcolor}
\usepackage{indentfirst}
\usepackage[numbers]{natbib}

\title{An ensemble solver for segregated cardiovascular FSI}

\author{
  Xue Li\\
  Department of Applied and Computational Mathematics and Statistics\\
  University of Notre Dame\\
  Notre Dame, IN 46556\\
  \texttt{xli26@nd.edu }\\
   \And
  Daniele E. Schiavazzi\\
  Department of Applied and Computational Mathematics and Statistics\\
  University of Notre Dame\\
  Notre Dame, IN 46556\\
  \texttt{dschiavazzi@nd.edu}\\
}

\begin{document}

\maketitle

\begin{abstract}
Computational models are increasingly used for diagnosis and treatment of cardiovascular disease. To provide a quantitative hemodynamic understanding that can be effectively used in the clinic, it is crucial to quantify the variability in the outputs from these models due to multiple sources of uncertainty. To quantify this variability, the analyst invariably needs to generate a large collection of high-fidelity model solutions, typically requiring a substantial computational effort.
In this paper, we show how an explicit-in-time \emph{ensemble} cardiovascular solver offers superior performance with respect to the embarrassingly parallel solution with implicit-in-time algorithms, typical of an inner-outer loop paradigm for non-intrusive uncertainty propagation.
We discuss in detail the numerics and efficient distributed implementation of a segregated FSI cardiovascular solver on both CPU and GPU systems, and demonstrate its applicability to idealized and patient-specific cardiovascular models, analyzed under steady and pulsatile flow conditions.
\end{abstract}

\section{Introduction}

\begin{figure}
\centering
\includegraphics[width=0.8\linewidth]{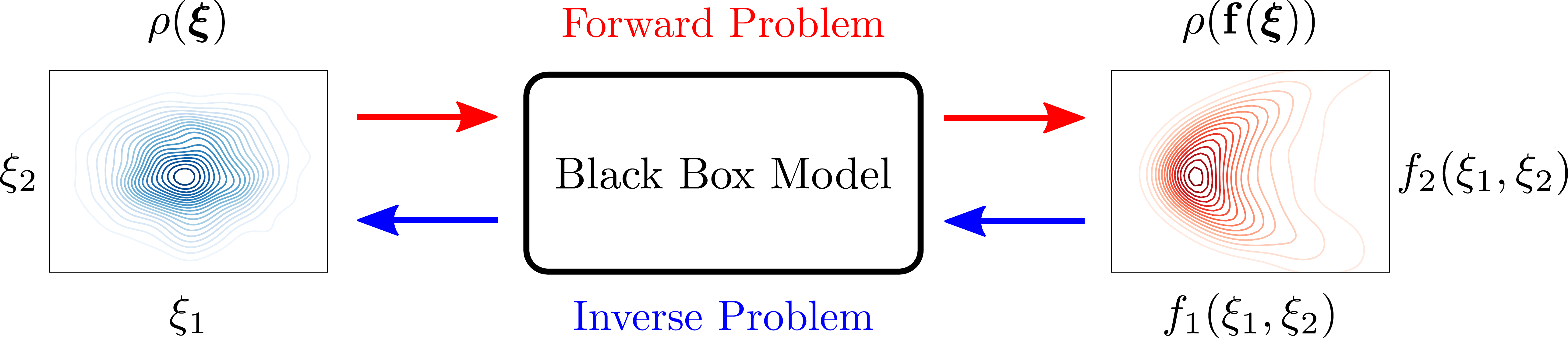}
\caption{Schematic representation of the forward and inverse problems in UQ.}\label{fig:uq}
\end{figure}

\noindent In this study we focus on the development of efficient solvers for complex fluid-structure interaction (FSI) phenomena arising in cardiovascular (CV) hemodynamics.
For this and many other applications, output variability is induced by uncertainty or ignorance in the input processes, e.g., material property distribution, physiologically sound boundary conditions or model anatomy, resulting from operator-dependent image volume segmentation. 
In this context, the new paradigm of Uncertainty Quantification (UQ) is rapidly becoming an integral part of the modeling exercise, and an indispensable tool to rigorously quantify confidence in the simulation outputs, enabling robust predictions of greater clinical impact.
However, running a complete UQ study on a large scale cardiovascular model is typically associated with a substantial computational cost.
Non-intrusive approaches for the solution of the forward problem in uncertainty quantification (also known as \emph{uncertainty propagation}) typically consider the underlying deterministic solver as a black box (see Figure~\ref{fig:uq}), requiring model solutions at various parameter realizations to be \emph{independently} (and possibly simultaneously) computed. 
The scalability of this paradigm is limited due to two main reasons. First, in the embarrassingly parallel solution of multiple instances of the same problem, a large number of operations is repeated. Second, the solution of large linear systems of equations from numerical integration in time with implicit schemes presents, in general, a less than ideal scalability when computed on large multi-core architectures.

To tackle these challenges, we propose an efficient computational approach for solving multiple instances of the same model (so-called model \emph{ensemble}) at the same time, in a highly scalable fashion, enabled by CPU/GPU implementations of numerical integration schemes that rely heavily on distributed sparse matrix-vector products.
We validate our approach by characterizing the effect of variability in vessel wall thickness and elastic modulus on the mechanical response of ideal and patient-specific cardiovascular models analyzed under steady and pulsatile flow conditions.
A stochastic model for the spatial distribution of thickness and elastic modulus is provided, in this study, by approximating a Gaussian random field with Mat\'ern covariance through the solution of a stochastic partial differential equation on a triangular finite element mesh of the vessel lumen~\cite{lindgren2011explicit}.
Additionally, we follow a one-way coupled, segregated approach to fluid-structure interaction, where the wall stress is computed by an implicit Variational Multiscale fluid solver and passed to a structural, three-d.o.f.s shell model of the vascular walls~\cite{figueroa2006coupled}.  
%
Unique contributions of our approach are:
\begin{enumerate}
\setlength\itemsep{0pt}
\item We propose, for the first time, an ensemble solver in the context of cardiovascular hemodynamics for UQ, with the aim of drastically reduce the computational effort to perform campaigns of high-fidelity model solutions.
\item Our approach paves the way to a fully explicit treatment of fluid structure interaction in cardiovascular modeling, whose potential has not yet been fully explored in the literature. In our opinion, this new paradigm will provide simpler approaches to simulating complex physiological dysfunctions (e.g. aortic dissection, involving large vessel deformations and contact/auto-contact phenomena, see~\cite{baumler2020fluid}).
\item The combination of explicit time integration schemes with the solution of model ensembles leads to an efficient distribution of computing load and memory usage in the GPU, enabling scalability in a way that is currently not possible with embarrassingly parallel model solutions and implicit solvers.
\end{enumerate}
We motivate the computational advantage of explicit-in-time ensemble solvers through a back-of-the-envelope argument. It is well known how explicit numerical integration schemes are only conditionally stable. For structural problems, the larger stable time step is determined through a CFL condition as $0.9\cdot\Delta t$ where $\Delta t$ is the amount of time required by an elastic wave to cross the smallest element in the mesh. 
A reasonable value of $\Delta t$ for CV modeling is $\Delta t = l_{e,\text{min}}/c$, where $l_{e,\text{min}}$ is the diameter of the circle/sphere inscribed in the smallest element in the mesh, $c = \sqrt{E/\rho}$ is the elastic wave speed, $E$ and $\rho$ the elastic modulus and density of the vascular tissue (assumed homogeneous and isotropic in the current argument).
Typical values of $l_{e,\text{min}} = 1.0\times 10^{-3}$ m, $E = 0.7$ MPa and $\rho = 1.06$ kg/m$^{3}$ lead to a stable time step equal to approximately $0.9\cdot 1.2\times$10$^{-6} \approx 1.0\times$10$^{-6}$.
In contrast, the typical time step adopted by implicit CV FSI solvers is one millisecond (an upper bound, typically much smaller). At every time step, these solvers require multiple linear systems to be solved with iterative methods, each consisting in several matrix-vector products per iteration. Thus, if we assume an average of 10 non-linear iterations consisting of 10 linear solver iterations each (with two matrix-vector multiplications per iteration), an explicit solver is only a factor of five more expensive. 
However, the cost of explicit methods can be further reduced through several means, for example, increasing the critical time step via mass scaling (see, e.g.,~\cite{askes2011increasing}), selectively updating the stiffness, damping and mass matrices between successive time steps to avoid the repeated assembly of element matrices (in the linear regime), and by solving multiple realizations of the boundary conditions, material properties and geometry at the same time.
Additionally, large matrix-vector operations needed by the explicit solution of model ensembles are particularly well suited for GPU computing. 
In summary, explicit time integration schemes have many advantages over implicit approaches in the simulation of phenomena occurring over small time intervals and, combined with the solution of model ensembles, bear substantial potential to boost the efficiency of solving CV models on modern GPU-based systems. 
Even though our work focuses on CV hemodynamics, the proposed solver paradigm is applicable to study fluid-structure interaction phenomena in other fields, but its efficiency is affected by the stiffness and mass properties in the selected application.

Use of explicit structural solvers in cardiovascular flow problems is mainly related to their flexibility in modeling complex contact configurations. Studies involving coronary stent deployment following endoscopic balloon inflation are proposed, e.g., in~\cite{morlacchi2011sequential,chiastra2020modeling}, and coupling with an implicit fluid solver is discussed in~\cite{chiastra2013computational}.
Implementation of structural explicit solvers on GPU are discussed in various studies in the literature. In~\cite{bartezzaghi2015explicit} the authors describe in detail an application involving thin shells, while an overview on applications in biomechanics is discussed in~\cite{strbac2017gpgpu}.
Additionally, Ensemble methods for fluid problems have been recently proposed by~\cite{jiang2014algorithm,jiang2015higher,jiang2015numerical,takhirov2016time,jiang2017second} in the context of the Navier-Stokes equations with distinct initial conditions and forcing terms. This is based on the observation that solution of linear systems is responsible for a significant fraction of the overall running time for linearly implicit methods, and that is far more efficient to solve multiple times a system of equation with the same coefficient matrix and different right-hand-side than different systems altogether. Extensions have also been proposed to magnetohydrodynamics~\cite{mohebujjaman2017efficient}, natural convection problems~\cite{fiordilino2018ensemble} and parametrized flow problems~\cite{luo2018ensemble,gunzburger2019efficient}.
We note how, in our case, there is no approximation introduced in the formulation of the ensemble numerical scheme. 
In addition, no ensemble method appears to be available from the literature in the context of fluid-structure interaction problems. 

The basic methodology behind the proposed solver is discussed in Section~\ref{sec:methods}, including the generation of random field material properties, the structural finite element formulation, the variational multiscale fluid solver, thier weak coupling and CPU/GPU implementation.
Validation of the proposed approach is discussed in Section~\ref{sec:results} with reference to an idealized model of the thoracic aorta and a patient-specific coronary model. Performance and scalability of the approach is discussed in Section~\ref{sec:solverPerformance} followed by conclusions and future work in Section~\ref{sec:discussion}.
 
\section{Methodology}\label{sec:methods}

\subsection{Random field material properties}\label{sec:randomField}

\noindent A homogeneous and isotropic vascular tissue with uncertain elastic modulus and thickness is assumed in this study, modeled through a Gaussian random field with Mat\'ern covariance. A Gaussian marginal distribution appears to be the simplest idealized distribution compatible with the scarce experimental observations, while the choice of a Mat\'ern covariance relates to its finite differentiability, which make this model more desirable than other kernels~\cite{stein2012interpolation}.
%
Let $\left\Vert\cdot\right\Vert$ denote the Euclidean distance in $\mathbb{R}^d$.
The Mat\'ern covariance between two points at distance \(\|\bm{h}\|\) is
\begin{equation}
    r(\|\bm{h}\|)=\frac{\sigma^2}{2^{\nu-1}\Gamma
(\nu)}(\kappa\|\bm{h}\|)^\nu K_\nu(\kappa\|\bm{h}\|),\,\,\bm{h}\in\mathbb{R}^d
\end{equation}
where $\Gamma(\cdot)$ is the gamma function, $K_\nu$ is the modified Bessel function of the second kind, $\sigma^2$ is the marginal variance, $\nu$ is a scaling parameter which determines the mean square differentiability of the underling process, and $\kappa$ is related to the correlation length $\rho=\sqrt{8\nu}/\kappa$, i.e., the distance which corresponds to a correlation of approximately $0.1$, for all $\nu$.
It is known from the literature~\cite{whittle1954stationary, whittle1963stochastic} how Gaussian random fields with Mat\'ern covariance can be obtained as solutions of a linear fractional stochastic partial differential equation (SPDE) of the form
\begin{equation}\label{eq:spde}
(\kappa^2-\Delta)^{\alpha/2}\,x(\bm{s})=\mathcal{W}(\bm{s}),\,\,\bm{s}\in\mathbb{R}^d,
\end{equation}
where $\alpha=\nu + d/2$, $\kappa>0$, $\nu>0$, $\mathcal{W}(\bm{s})$ is a white noise spatial process, $\Delta=\Sigma_i\,\partial^2 / \partial\,s_i^2$, and the marginal variance is
\[
\sigma^2=\frac{\Gamma(\nu)}{\Gamma(\nu+d/2)(4\pi)^{d/2}\kappa^{2\nu}}.
\]

Since the lumen wall is modelled with a surface of triangular elements, we are interested in generating realizations from discretely indexed Gaussian random fields. This is achieved through an approximate stochastic weak solution of the SPDE~\eqref{eq:spde}, as discussed in~\cite{lindgren2011explicit}.
We construct a discrete approximation of the solution $x(\bm{s})$ using a linear combination of basis functions, $\{\psi_k\},k=1,\dots,n$, and appropriate weights, $\{w_k\},k=1,\dots,n$, i.e., $x(\bm{s})=\sum_{k=1}^{n}\,\psi_k(\bm{s})\,w_k$. 
We then introduce an appropriate Sobolev space with inner product $\langle\cdot,\cdot\rangle$, a family of \emph{test functions} $\{\varphi_{k}\},k=1,\dots,n$, and derive a Galerkin functional for~\eqref{eq:spde} of the form 
\begin{equation}
\langle\varphi_i,(\kappa^2-\Delta)^{\alpha/2}\,\psi_j\rangle\,w_{j} =\langle\varphi_i,\mathcal{W}\rangle
\end{equation}
We then choose $\varphi_{k} = (\kappa^{2} - \Delta)^{1/2}\,\psi_{k}$ for $\alpha=1$ and $\varphi_{k} = \psi_{k}$ for $\alpha=2$, leading to precision matrices $\bm{Q}_{\alpha}$ expressed as 
\begin{equation}
\begin{aligned}
&\bm{Q}_{\alpha}=\varkappa^2\bm{C+G} && \text{for}\,\,\alpha=1\\
&\bm{Q}_{\alpha}=(\varkappa^2\bm{C+G})^T\bm{C}^{-1}(\varkappa^2\bm{C+G}) && \text{for}\,\,\alpha=2\\
&\bm{Q}_\alpha=(\varkappa^2\bm{C+G})^T\bm{C}^{-1}\bm{Q}_{\alpha-2}\bm{C}^{-1}(\varkappa^2\bm{C+G}) && \text{for}\,\,\alpha > 2,
\end{aligned}
\end{equation}
where, for $\alpha \ge 3$ a recursive Galerkin formulation is used, letting $\alpha=2$ on the left-hand side of equation~\eqref{eq:spde} and replacing the right-hand side with a field generated by $\alpha-2$, assigning $\varphi_{k} = \psi_{k},\,k=1,\dots,n$.
Note how the use of piecewise linear basis functions $\{\psi_{k}\},k=1,\dots,n$, lead to matrices
\begin{equation}
\bm{G}_{ij}=\langle\nabla\psi_i,\nabla\psi_j\rangle\,\,\text{and}\,\,\bm{C}_{ij}=\langle\psi_i,\psi_j\rangle,
\end{equation}
that are \emph{sparse}, and often found in the finite element discretization of second order elliptic PDEs.
However, the precision matrices $\bm{Q}_{\alpha}$ are, in general, not sparse as they contain the inverse $\bm{C}^{-1}$. Thus, by replacing the matrix $\bm{C}$ with the \emph{lumped} diagonal matrix $\widetilde{\bm{C}}$, sparsity is restored and $\bm{Q}_{\alpha}$ can be efficiently manipulated and decomposed through fast routines for sparse linear algebra available on a wide range of architectures. 
In addition, the introduction of $\widetilde{\bm{C}}$ leads to non zero terms on each row only for the \emph{immediate neighbors} $\mathcal{I}(s_{k})$ of a given node $k$ on the triangular surface mesh, since the basis function $\{\psi_{k}\}$ is supported only on the elements connected to node $k$. This reduces the Gaussian random field to a Gaussian Markov random field, for which 
\begin{equation}\label{equ:markov}
\begin{split}
& \rho\left(x(s_{i}),x(s_{k})\,\vert\,x(s_{j}),\,s_{j}\in\mathcal{I}(s_{i}) \right) = \\
& =\rho\left(x(s_{i})\,\vert\,x(s_{j}),\,s_{j}\in\mathcal{I}(s_{i}) \right),
\end{split}
\end{equation}
or, in other words, \emph{given the value of $x$ on its neighbors}, at any node $k$ the random field $x(s_{k})$ is statistically independent from any other location. The approximation error introduced in~\eqref{equ:markov} is, however, small and often negligible in applications~\cite{bolin2013comparison}. The interested reader is referred to~\cite{lindgren2011explicit} for additional detail on the derivation of $\bm{Q}_{\alpha}$.

\begin{figure}[!ht]
\centering
\begin{minipage}{0.45\textwidth}
\begin{subfigure}[b]{0.3\textwidth}
\centering
\includegraphics[width=0.73\textwidth,trim={1.3cm 0cm 0.7cm 0cm},clip]{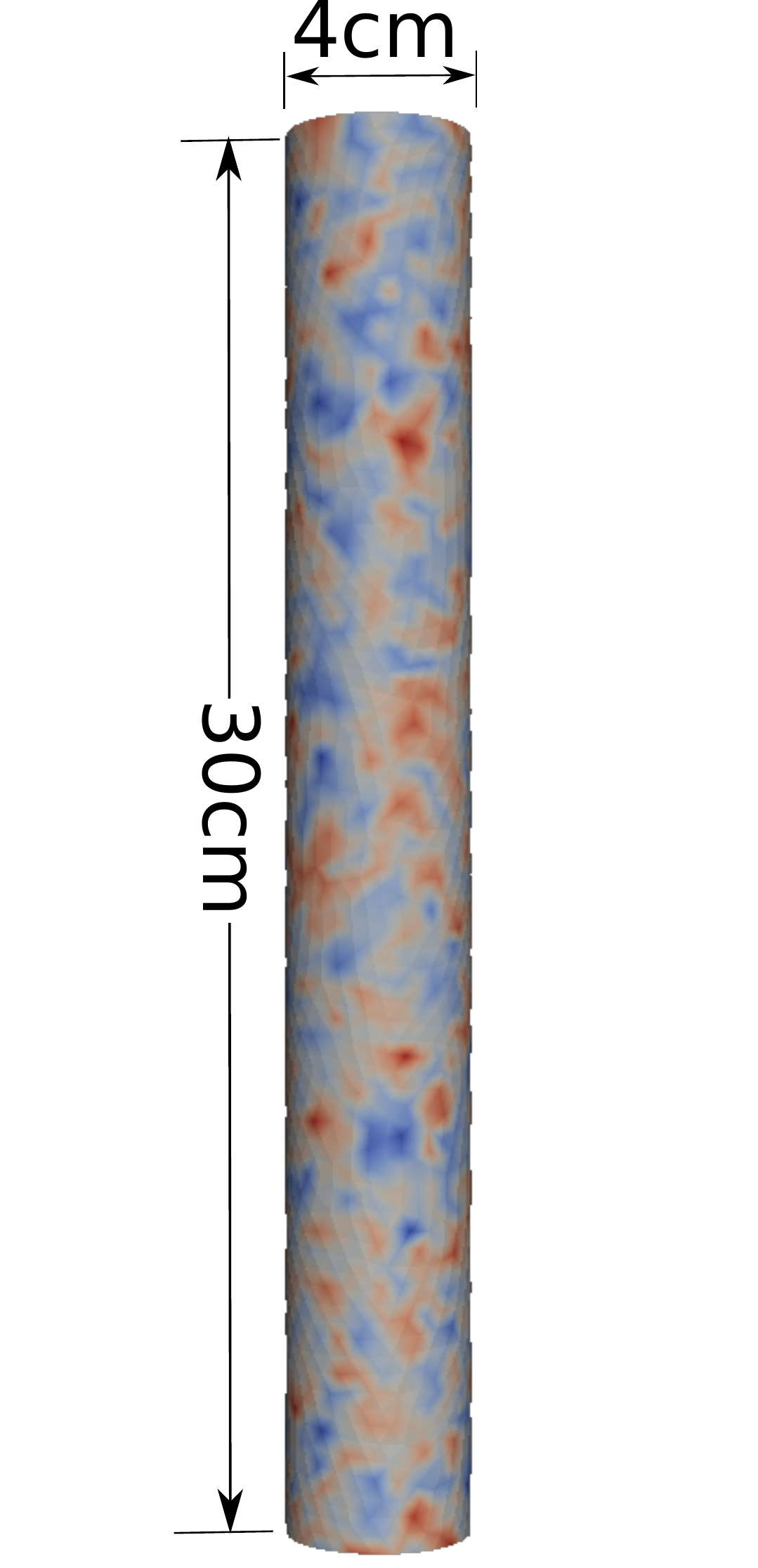}
\caption{$\rho=.95$\,cm}
\end{subfigure}
\begin{subfigure}[b]{0.3\textwidth}
\centering
\includegraphics[width=0.86\textwidth,trim={1.5cm 0cm 1.5cm 0cm},clip]{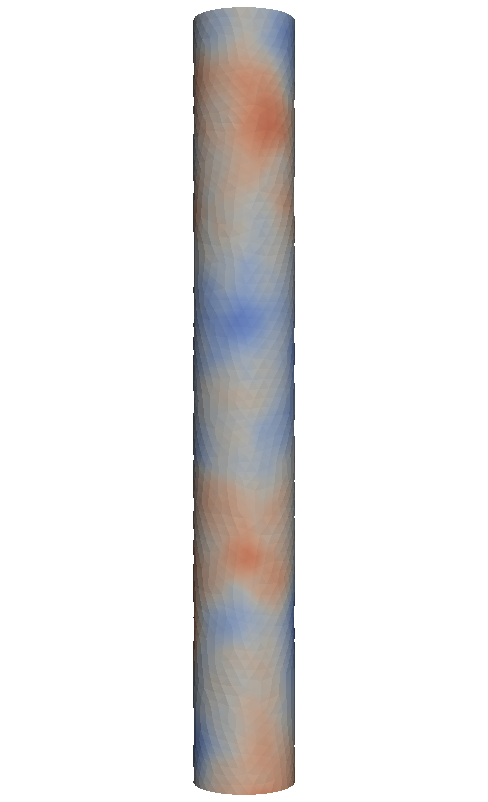}
\caption{$\rho=3.7$\,cm}
\end{subfigure}
\begin{subfigure}[b]{0.3\textwidth}
\centering
\includegraphics[width=0.9\textwidth,trim={1.5cm 0cm 1.5cm 0cm},clip]{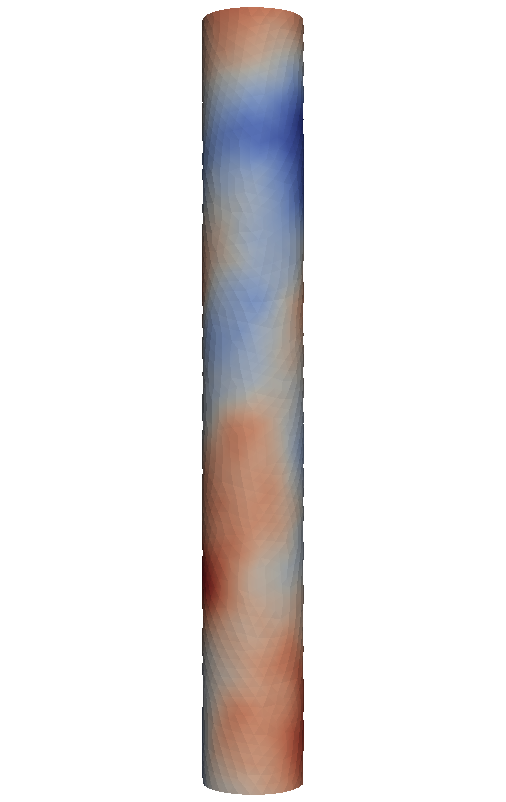}
\caption{$\rho=7.2$\,cm}
\end{subfigure}
\caption{Random field generated on a cylindrical mesh for various correlation lengths.}\label{fig:04}
\end{minipage}
\hspace{3pt}
\begin{minipage}{0.45\textwidth}
\centering
\includegraphics[width=0.9\textwidth,trim={0cm 0.3cm 0cm 1.3cm},clip]{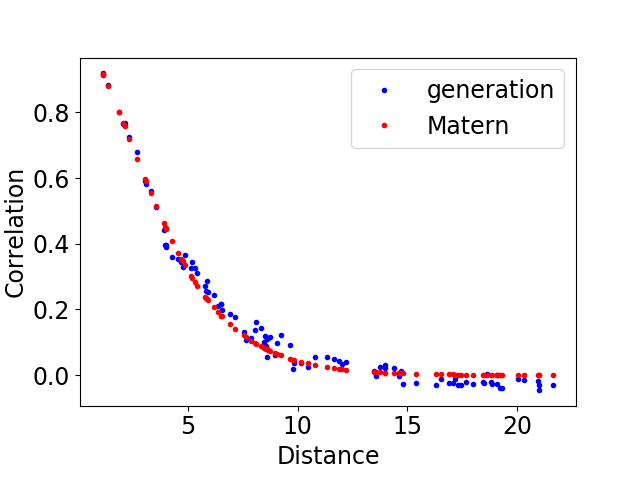}
\caption{Comparison between spatial correlations from a  numerically generated field $x(\bm{s})$ with precision matrix $\bm{Q}_{\alpha}$ and the exact Mat\'ern model.}\label{fig:05}
\end{minipage}
\end{figure}

Numerically generated realizations for various correlation lengths are shown in Figure~\ref{fig:04} on an ideal cylindrical representation of the descending thoracic aorta, while Figure~\ref{fig:05} shows the agreement of the generated field with the Mat\'ern model. 

\subsection{A segregated solver for fluid-structure interaction phenomena}\label{sec:segregated}

\subsubsection{Finite element model for the vessel wall}\label{sec:structFE}

\noindent We use a small strain, linear, 3 d.o.f. elastic thin shell which allows for a full compatibility between the fluid mesh discretized with tetrahedral elements and the solid walls.
The in-plane stiffness of the shell is complemented with a transverse shear stiffness which provides stability under transverse loading~\cite{figueroa2006coupled}.
%
Using a superscript $l$ and lower case $x,y,z$ to indicate quantities expressed in the local shell reference frame, we introduce a constitutive relation in Voigt notation expressed as
\begin{equation}
\setlength\arraycolsep{2pt}
\boldsymbol{\sigma}^{l}=\bm{C}\cdot\boldsymbol{\varepsilon}^{l},\,\,\text{with}\,\,
\boldsymbol{\sigma}^{l}=
\begin{bmatrix}
\sigma_{xx}\\
\sigma_{yy}\\
\tau_{xy}\\
\tau_{xz}\\
\tau_{yz}
\end{bmatrix},
\boldsymbol{\varepsilon}^{l}=
\begin{bmatrix}
\partial u_{x} / \partial x\\ 
\partial u_{y} / \partial y\\ 
\left(\partial u_{x} / \partial y + \partial u_{y} / \partial x \right)\\
\partial u_{z} / \partial x\\
\partial u_{z} / \partial y
\end{bmatrix}.
\end{equation}
We assume $\varepsilon^{l}_{zz}=0$, i.e., zero deformation through the thickness, disregarding the effect of both the normal pressure acting at the lumen surface and the Poisson effect due to the membrane deformations.
%
%
Strains $\boldsymbol{\varepsilon}^l$ and nodal displacements $\bm{u}$ are related through the matrix $\bm{B}$ of shape function derivatives for a linear triangular element, i.e.
\begin{equation}
\boldsymbol{\varepsilon}^l=\bm{B}\,\bm{u} = 
\frac{1}{2\,A_e}
\begin{bmatrix}
y_{23} & 0 & 0 & y_{31} & 0 & 0 & y_{12} & 0 & 0\\
0 & x_{32} & 0 & 0 & x_{13} & 0 & 0 & x_{21} & 0\\
x_{32} & y_{23} & 0 & x_{13} & y_{31} & 0 & x_{21} & y_{12} & 0\\
0 & 0 & y_{23} & 0 & 0 & y_{31} & 0 & 0 & y_{12}\\
0 & 0 & x_{32} & 0 & 0 & x_{13} & 0 & 0 & x_{21}
\end{bmatrix}
\begin{bmatrix}
u_{x,1}\\
u_{y,1}\\
u_{z,1}\\
u_{x,2}\\
u_{y,2}\\
u_{z,2}\\
u_{x,3}\\
u_{y,3}\\
u_{z,3}
\end{bmatrix}
\end{equation}
where $x_{j},y_{j},\,j\in\{1,2,3\}$ are the local coordinates of the $j$-th element node, $x_{ij}=x_i-x_j$ (similarly for $y_{ij}$), $u_{x,j},u_{y,j},u_{z,j}$ are the local nodal displacements and $A_e$ is the triangular element area.
The constitutive matrix is expressed as 
\begin{equation}
\bm{C}=\frac{E}{(1-\nu^2)}
\begin{bmatrix}
1 & \nu & 0 & 0 & 0\\
\nu & 1 & 0 & 0 & 0\\
0 & 0 & 0.5\,(1-\nu) & 0 & 0\\
0 & 0 & 0 & 0.5\,k\,(1-\nu) & 0\\
0 & 0 & 0 & 0 & 0.5\,k\,(1-\nu)
\end{bmatrix}
\end{equation}
where $E$ and $\nu$ are the Young's modulus and Poisson's ratio coefficient, respectively, and the shear factor $k$ accounts for a parabolic variation of transverse shear stress through the shell thickness (assumed as $5/6$ for a shell with rectangular cross section).
Finally, the local element stiffness matrix $\bm{k}_{e}\in\mathbb{R}^{9\times 9}$ can be expressed as
\begin{equation}
\bm{k}_{e}=\int_{\Omega^{s}_e}\,\bm{B}^T\bm{C}\bm{B}\,\,\mathrm{d}\Omega^{s}_{e} = \sum_{i=1}^{n_{\text{gp}}}\,\bm{B}^T\bm{C}\bm{B}\,A_e\,\zeta_{i}\,w_{i},
\end{equation}
where $n_{\text{gp}}$ is the total number of integration points and $\zeta_{i},w_{i},i\in\{1,2,\dots,n_{\text{gp}}\}$, are the element thickness and integration rule weights, respectively.
In this study, we adopt a three-point Gauss integration rule to capture a linear variation for $E$ and $\zeta$ through each element, generated from a Gauss Markov random field with Mat\'ern covariance $\bm{Q}_{\alpha}$, i.e., $E=E(x,y,\omega)$ and $\zeta = \zeta(x,y,\omega)$. Nodal vectors $\bm{E}\sim\mathcal{N}(\overline{\bm{E}},\bm{Q}^{-1}_{\alpha})$ and $\bm{\zeta}\sim\mathcal{N}(\overline{\bm{\zeta}},\bm{Q}^{-1}_{\alpha})$ are generated as
\begin{equation}
\bm{E}=\overline{\bm{E}}+(\bm{L}^T)^{-1}\bm{z},\,\,\text{and}\,\,\bm{\zeta}=\overline{\bm{\zeta}}+(\bm{L}^T)^{-1}\bm{z},
\end{equation}
where $\bm{z}\sim\mathcal{N}(\bm{0},\bm{I}_{n})$ is a vector with standard Gaussian components and $\bm{L}$ is the sparse Cholesky factor of $\bm{Q}_{\alpha}$, i.e., $\bm{Q}_{\alpha} = \bm{L}\,\bm{L}^{T}$. 
The covariance matrix $\bm{Q}_{\alpha}$ is assembled in compressed sparse column (CSC) format and the Cholesky decomposition is computed using the \emph{cholmod} routine provided by the \emph{scikit-sparse} Python library. Finally, the product $(\bm{L}^T)^{-1}\bm{z}$ is performed using the \emph{solve\_Lt} routine, as the solution of a triangular system.

\subsubsection{Variational multiscale finite element fluid solver}\label{sec:VMS}

\noindent The evolution of blood flow and pressure in the human cardiovascular system can be modeled using the Navier-Stokes equations. Even though many simplifying assumptions can be made to the equations to reduce the computational complexity, here we focus on high-fidelity models, i.e., models associated with large discretizations of a three-dimensional ($n_{\text{sd}}=3$) fluid domain $\Omega^{f}\subseteq\mathbb{R}^{n_{\text{sd}}}$.
The boundary $\Gamma^{f}$ of $\Omega^{f}$ coincides with the mid-plane of the solid domain $\Omega^{s}$, and is partitioned into $\Gamma^{f} = \Gamma_{g}^{f}\cup\Gamma_{h}^{f}\cup\Gamma_{s}^{f}$ which correspond to the application of Dirichlet, Neumann boundary conditions and interaction with the solid, respectively. 
Consider also the vector fields $\bm{h}:\Gamma_h\times(0,T)\to\mathbb{R}^{n_{sd}}$, $\bm{g}:\Gamma_g\times(0,T)\to\mathbb{R}^{n_{sd}}$, $\bm{f}:\Omega\times(0,T)\to\mathbb{R}^{n_{sd}}$ and $\bm{v}^{0}:\Omega\to\mathbb{R}^{n_{sd}}$.
We would like to solve the problem of finding $\bm{v}(\bm{X},t)$ and $p(\bm{X},t)$, $\forall\,\bm{X}\in\Omega$, $\forall\,t\in[0,T]$ such that
\begin{equation}\label{equ:strongNS}
\begin{cases} 
\rho\,\dot{\bm{v}} + \rho\,\bm{v}\cdot\nabla\bm{v} = -\nabla p + \nabla\cdot\boldsymbol{\tau}+ \bm{f}  & (\bm{X},t)\in \Omega\times(0,T)\\
\nabla\cdot\bm{v}=0 & (\bm{X},t)\in \Omega\times(0,T),\\
\end{cases}
\end{equation}
subject to the boundary conditions
\begin{equation}
\begin{cases}
\bm{v}=\bm{g}  & (\bm{X},t)\in\Gamma_g\times(0,T)\\
\bm{t}_{\bm{n}} = \bm{\sigma}\cdot\bm{n}=[-p\,\bm{I}+\bm{\tau}]\cdot\bm{n}=\bm{h} & (\bm{X},t)\in\Gamma_h\times(0,T)\\
\bm{t}_{\bm{n}}=\bm{t}^f & (\bm{X},t)\in\Gamma_s\times(0,T)\\
\bm{v}(\bm{X},0)=\bm{v}^{0}(\bm{X}) & \bm{X}\in\Omega,
\end{cases}
\end{equation}
where $\bm{\tau}=\mu(\nabla\bm{v}+\nabla\bm{v}^{\,T})$ is the viscous stress tensor resulting from considering blood as a Newtonian fluid. 
Solution of \eqref{equ:strongNS} in weak form requires to define four approximation spaces, i.e., two trial spaces for the velocity $\bm{v}$ and pressure $p$
\begin{equation*}
\begin{split}
\mathscr{S}_k^h & = \Big\{\bm{v}\,\vert\,\bm{v}(\cdot,t)\in\bm{H}^1(\Omega),\mathnormal{t}\in[0,\mathnormal{T}],\\
& \bm{v}\,\vert_{\bm{x}\in\Omega_{e}}\in\mathnormal{P}_k(\Omega_{e}),\,\bm{v}(\cdot,t)=\bm{g} \text{ on } \Gamma_g\Big\},\\
\mathscr{P}_k^h & =\left\{p\,\vert\,p(\cdot,t)\in L^{2}(\Omega),t\in[0,\mathnormal{T}],\,p\,\vert_{\bm{x}\in\bar{\Omega}_e}\in\mathnormal{P}_k(\Omega_e)\right\},
\end{split}
\end{equation*}
and two test spaces for $\vec{w}$ and $q$
\begin{equation*}
\begin{split}
\mathscr{W}_k^h = \Big\{ & \bm{w}\,\vert\,\bm{w}(\cdot,t)\in\bm{H}^1(\Omega),\,\mathnormal{t}\in[0,\mathnormal{T}],\\
& \bm{w}\,\vert_{\bm{x}\in\Omega_{e}}\in\mathnormal{P}_k(\Omega_{e}),\,\bm{w}(\cdot,t)=\bm{0} \text{ on } \Gamma_g\Big\},\,\,\mathscr{Q}_k^h = \mathscr{P}_k^h
\end{split}
\end{equation*}
where $\bm{H}^{1}(\Omega)$ is the Sobolev space of function triplets in $L^{2}(\Omega)$ with derivatives in $L^{2}(\Omega)$ and $P_{k}(\Omega)$ is the space of polynomials of order $k$ in $\Omega$.
%
The above spaces are separated into large and a small scale contributions $\mathscr{S}_k^h = \overline{\mathscr{S}_k^h} \oplus \widetilde{\mathscr{S}_k^h}$, $\mathscr{W}_k^h = \overline{\mathscr{W}_k^h} \oplus \widetilde{\mathscr{W}_k^h}$ and $\mathscr{P}_k^h = \overline{\mathscr{P}_k^h}\oplus \widetilde{\mathscr{P}_k^h}$ with a corresponding decomposition of velocity and pressures as $\bm{v} = \overline{\bm{v}} + \widetilde{\bm{v}}$ and $p = \overline{p} + \widetilde{p}$, respectively.
This decomposition is introduced in a weak form of the Navier-Stokes equations~\eqref{equ:strongNS} and a \emph{closure} obtained by expressing the small scale variables $\widetilde{\bm{v}}$ and $\widetilde{p}$ in terms of their large scale counterparts  $\overline{\bm{v}}$ and $\overline{p}$ using~\cite{bazilevs2007variational}
\begin{equation}
\begin{bmatrix}
\widetilde{\bm{v}}\\
\widetilde{\bm{p}}
\end{bmatrix} = 
\begin{bmatrix}
\tau_{M}\,\bm{R}_{M}(\overline{\bm{v}}, \overline{p})\\
\tau_{C}\, R_{C}(\overline{\bm{v}}, \overline{p})
\end{bmatrix},
\end{equation}
where $\bm{R}_{M}$ and $R_{C}$ represent the momentum and continuity residual expressed as
\begin{equation}
\begin{cases}
\bm{R}_{M}(\overline{\bm{v}},\overline{p}) = \bm{R}_{M} =\rho\,\dot{\overline{\bm{v}}} + \rho\,\overline{\bm{v}}\cdot\nabla\overline{\bm{v}}+\nabla\overline{p}-\nabla\cdot\bm{\tau}-\bm{f},\\
R_{C}(\overline{\bm{v}},\overline{p}) = R_{C} = \nabla\cdot\overline{\bm{v}},
\end{cases}
\end{equation}
and the \emph{stabilization} coefficients are expressed as
\begin{equation}
\begin{split}
\tau_{M} & = \left(\frac{4}{\Delta t^{2}} + \bm{u}\cdot\bm{G}\,\bm{u} + C_{I}\,\nu^{2}\,\bm{G}:\bm{G}\right)^{-1/2}\\
\tau_{C} & = (\tau_{M}\,\bm{g}\cdot\bm{g})^{-1},\,\,G_{i,j} = \sum_{k=1}^{3}\,\dfrac{\partial \xi_{k}}{\partial x_{i}}\,\dfrac{\partial \xi_{k}}{\partial x_{j}},\,\,\,g_{i} = \sum_{k=1}^{3}\,\dfrac{\partial \xi_{k}}{\partial x_{i}},
\end{split}
\end{equation}
where $\partial \bm{\xi} / \partial \bm{x}$ is the inverse Jacobian of the element mapping between the parametric and physical domains.

To simplify the notation, in what follows the large scale variables $\overline{\bm{v}}$ and $\overline{p}$ will be denoted simply by $\bm{v}$ and $p$ and the spaces $\overline{\mathscr{S}_k^h}$, $\overline{\mathscr{W}_k^h}$ and $\overline{\mathscr{P}_k^h}$ by $\mathscr{S}_k^h$, $\mathscr{W}_k^h$ and $\mathscr{P}_k^h$.
A weak solution of the Navier-Stokes equations can now be determined by finding $\bm{v}\in\mathscr{S}_h^k$ and $p\in\mathscr{P}_h^k$ such that
\begin{equation}\label{equ:weakNS}
\begin{split}
& B(\bm{w},q\,;\,\bm{v},p)= B_G(\bm{w},q\,;\,\bm{v},p) +\\
&+\sum_{e=1}^{n_{\text{el}}}\,\int_{\Omega_e}\left\{(\bm{v}\cdot\nabla)\,\bm{w}\cdot(\tau_M\,\bm{R}_{M})+\nabla\cdot\bm{w}\,\tau_C\,R_{C}\right\}\,\mathrm{d}\Omega_{e} +\\
&+\sum_{e=1}^{n_{\text{el}}}\,\int_{\Omega_e}\left\{\bm{w}\cdot\left[-\tau_M\,\bm{R}_{M}\cdot\nabla\bm{v}\right]+\left[\bm{R}_{M}\cdot\nabla\bm{w}\right]\cdot\left[\overline{\tau}\,\bm{R}_{M}\cdot\bm{v}\right]\right\}\,\mathrm{d}\Omega_{e} +\\
&+\sum_{e=1}^{n_{\text{el}}}\,\int_{\Omega_{e}}\nabla q\cdot\frac{\tau_M}{\rho}\,\bm{R}_{M}\,\mathrm{d}\Omega_{e}=0,
\end{split}
\end{equation}
for all $\vec{w}\in\vec{\mathscr{W}}_h^k$ and $q\in\mathscr{P}_h^k$, where the Galerkin functional $B_G$ is expressed as
\begin{equation}
\begin{split}
& B_G(\bm{w},q\,;\,\bm{v},p)=\int_{\Omega} \bm{w}\cdot(\rho\,\dot{\bm{v}}+\rho\,\bm{v}\cdot\nabla\bm{v}-\bm{f})\,\,\mathrm{d}\Omega + \\
& + \int_{\Omega} \left\{\nabla\bm{w}\,:\,(-p\bm{I}+\bm{\tau})-\nabla q\cdot\bm{v}\right\}\,\mathrm{d}\Omega +\\
& +\int_{\Gamma_h}\,\left\{-\bm{w}\cdot\bm{h}+q\,v_n\right\}\,\mathrm{d}\Gamma+\int_{\Gamma_s}\,\left\{-\bm{w}\cdot\bm{t}^f+q\,v_n\right\}\,\mathrm{d}\Gamma +\\
& + \int_{\Gamma_g}\,q\,v_n\,\mathrm{d}\Gamma.
\end{split}
\end{equation}
%

The discrete variables $\bm{w}^{h}$, $\bm{v}^{h}$, $q^{h}$ and $p^{h}$ are introduced in~\eqref{equ:weakNS}, leading to a non linear system of equations of the form
\begin{equation}\label{equ:nonlinEq}
\begin{cases}
\bm{N}_{M}(\dot{\bm{v}}^{h},\bm{v}^{h},p^{h}) = 0\\
N_{C}(\dot{\bm{v}}^{h},\bm{v}^{h},p^{h}) = 0,
\end{cases}
\end{equation}
A predictor-multicorrector scheme~\cite{jansen2000generalized} is used for time integration and, at each time step, the resulting non linear system~\eqref{equ:nonlinEq} is solved through successive Newton iterations.
Additional details can be found in~\cite{bazilevs2007variational,seo2019performance}.
%
%

\subsubsection{Fluid-structure coupling}\label{sec:FSI}

\begin{figure}[!ht]
\centering
\includegraphics[width=0.5\linewidth]{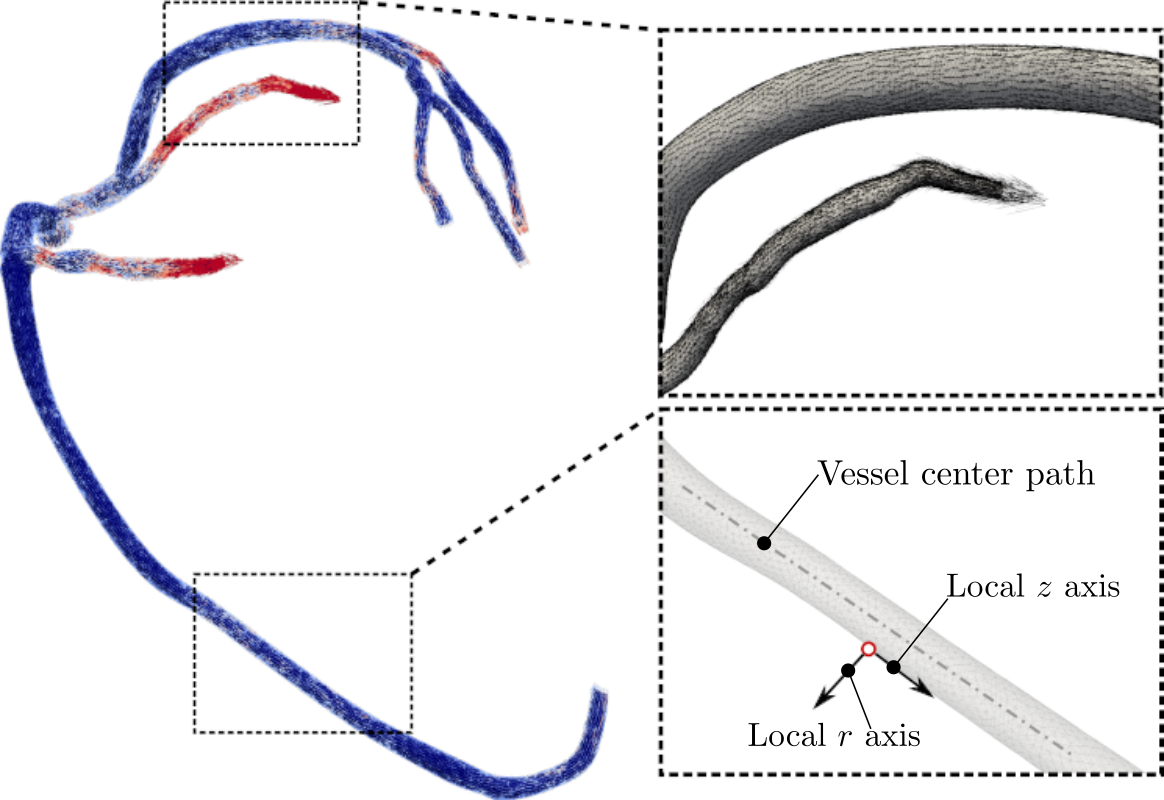}
\caption{Visualization of interface shear forces exchanged between the fluid and the structural solver. The local axis system used for stress post-processing is also shown in the close up.}\label{fig:interfaceShear}
\end{figure}

\noindent Since this study focuses more on the development of an ensemble solver, we provide a one-way coupled, simplified treatment of the interaction between fluid and structure, leaving a more rigorous treatment to future work. Here we assume a fluid model with rigid walls and a structural model of the lumen governed by the three d.o.f. shell discussed in Section~\ref{sec:structFE}. The lumen deformation does not, in turn, affect the geometry of the fluid region. 
The elastic forces in the lumen wall are in equilibrium with the shear and pressure exerted by the fluid, or, in other words $\bm{t}^s=-\bm{t}^f$, where the wall stress $\bm{t}^f$ computed by the fluid solver is
\begin{equation}
\bm{t}^f=\bm{\sigma}^{f}\cdot\bm{n},\,\,\bm{\sigma}=2\mu\bm{\varepsilon}-p\bm{I},
\end{equation}
$p$ is the main \emph{nodal} pressure unknown in the VMS fluid solver, and $\bm{\varepsilon} = \nabla^{s}\bm{u}$ is the symmetric part of the velocity gradient, which is constant on each P1 element in the fluid mesh. An example of shear forces computed by the VMS fluid solver for a coronary model is illustrated in Figure~\ref{fig:interfaceShear}.
At every node on the wall, the shear forces and normal vectors from adjacent elements are averaged and the nodal pressure added, leading to the three components of the nodal force that are passed to the structural solver.

Finally, stress components in the circumferential, radial and axial directions are obtained by transforming the solid stress tensor to a local cylindrical coordinate system (see Figure~\ref{fig:interfaceShear}).
For an arbitrary Gauss point or lumen shell node $\bm{s}$, we identify the closest location on the vessel centerline. The tangent vector to the centerline at $\bm{s}$ is defined as the local \emph{axial} direction $z$, the normal at the Gauss point is the local \emph{radial} direction $r$, and the local \emph{circumferential} direction $\theta$ is obtained as the cross product between $r$ and $z$.

\subsection{Distributed explicit FSI solver on multiple CPUs}\label{sec:fsiCPU}

\noindent The solution in time for the dynamics of the undamped non-linear structural system 
%
\begin{equation}
\bm{M}\,\bm{\ddot{u}} + \bm{C}\,\bm{\dot{u}} + \bm{K}(\bm{u})\,\bm{u} = \bm{f}(t),
\end{equation}
%
is computed using central differences in time, resulting in an update formula in the $[n\,\Delta t,(n+1)\,\Delta t]$ intervals expressed as
\begin{equation}\label{equ:dynamicsCD}
\left(\bm{\widetilde{M}}+\frac{\Delta t}{2}\bm{\widetilde{C}}\right)\,\boldsymbol{u}_{n+1}=\Delta t^2\,\bm{f}_{n}-
\left(\Delta t^2\bm{K}-2\,\bm{M}\right)\,\boldsymbol{u}_n-\left(\bm{M}-\frac{\Delta t}{2}\bm{C}\right)\,\boldsymbol{u}_{n-1}
\end{equation}
which is performed independently by an arbitrary number of mesh partitions.
To do so efficiently, 
$\bm{\widetilde{M}},\bm{\widetilde{C}}$ in~\eqref{equ:dynamicsCD} are lumped mass matrix
pre-assembled before the beginning of the time loop, containing the elemental contributions from all finite elements, even those belonging to separate mesh partitions. Given the limited amount of deformation typically observed in cardiovascular applications over a single heart cycle, the assumption of a fixed nodal mass over time is considered realistic. This assumption removes the need of communicating nodal masses during finite element assembly, improving scalability.
In some cases, a viscous force $\bm{f}_{v}$ is added to the right-hand-side in order to damp the high-frequency oscillations as $\bm{f}_{v} = -c_{d}\,\dot{\bm{u}}_{n}$, through an appropriate damping coefficient $c_{d}$.

Even though the geometry of the vessel lumen is updated at every step by adding $\bm{u}_{n+1}$, the displacement magnitudes over a typical heart cycle remain of the same order as the thickness of the vessel wall, hence in the linear regime. In this context, the stiffness do not significantly change and it is possible to save computational time by avoiding to assemble it at every iteration. In our code, we therefore provide the option to selectively update the stiffness matrix after a prescribed number of iterations.
Synchronization of displacements for the nodes shared by multiple partitions is implemented using \emph{Send}, \emph{Recv} to the root CPU and broadcast back.

\begin{figure*}[!ht]
\centering
\begin{subfigure}[b]{0.32\linewidth}
\centering
\includegraphics[width=\linewidth]{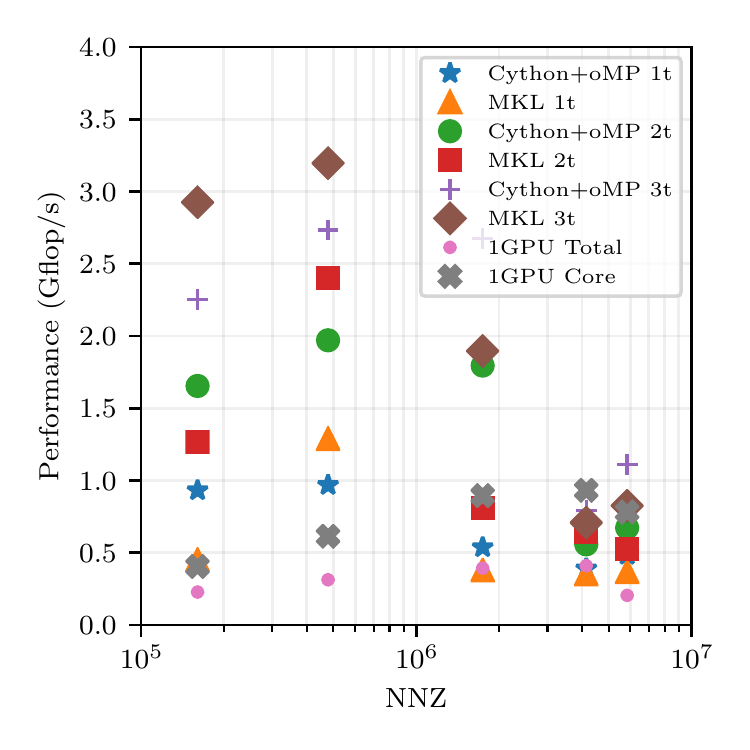}
\caption{}
\end{subfigure}
\begin{subfigure}[b]{0.41\linewidth}
\centering
\includegraphics[width=\linewidth]{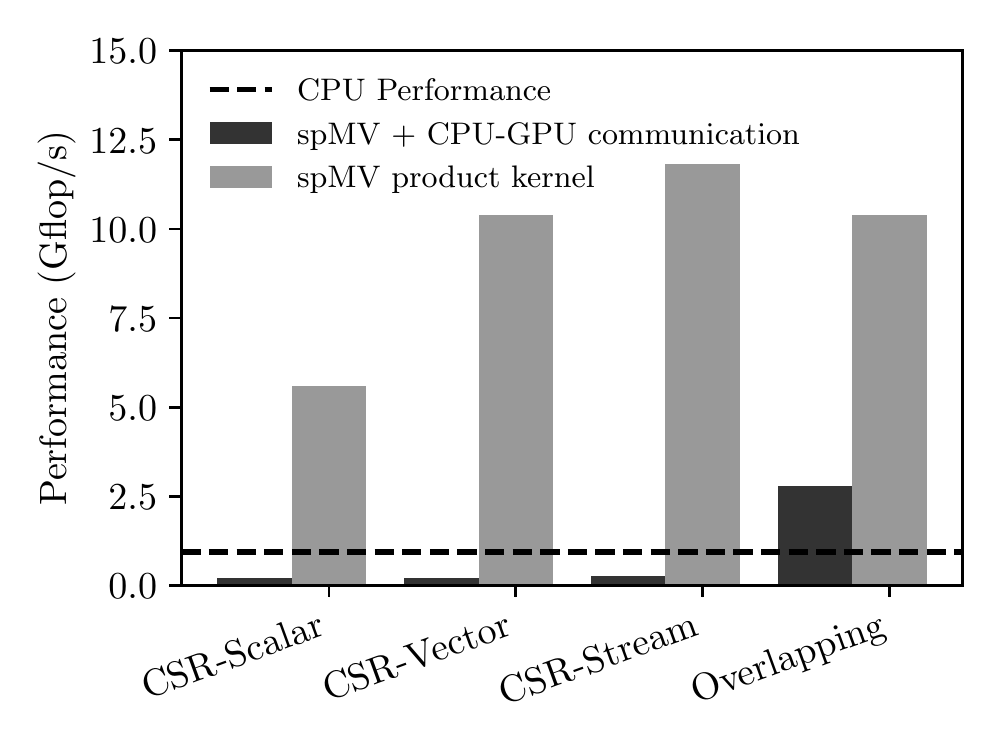}
\caption{}\label{fig:gpuDataTransfer}
\end{subfigure}
\caption{Performance of matrix-vector product kernel on CPU by our Cython+openMP implementation, GPU implementation and MKL library on multiple threads (a). Optimization of GPU matrix-vector kernel and CPU-GPU communication performance (b).
Tests were performed using 1 CPU and 1 GPU on a cylindrical model associated with a sparse matrix having nnz = 160,587 and 1,000 material property realizations.}\label{fig:matProd}
\end{figure*}

\begin{figure*}[!ht]
\centering
\begin{subfigure}[b]{0.32\linewidth}
\centering
\includegraphics[width=\linewidth]{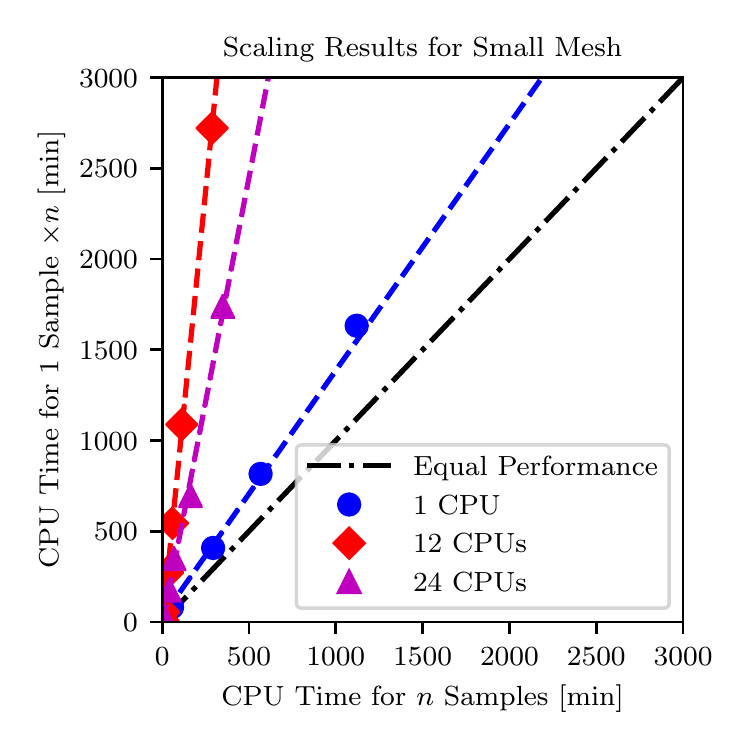}
\caption{}
\end{subfigure}
\begin{subfigure}[b]{0.32\linewidth}
\centering
\includegraphics[width=\linewidth]{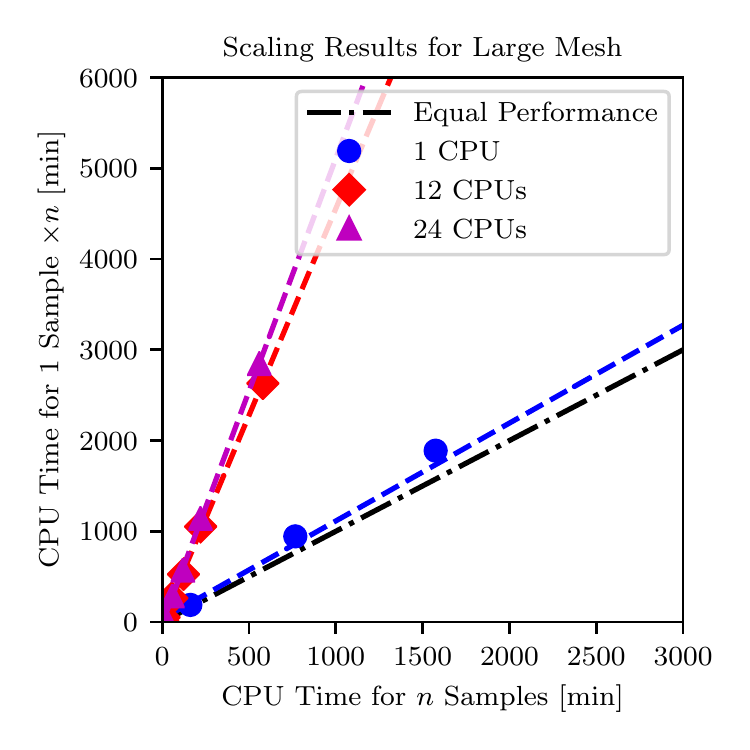}
\caption{}
\end{subfigure}
\caption{Performance of explicit ensemble solver on multiple CPUs for a mesh with 5,074 elements, 2,565 nodes (a) and for a mesh with 15,136 elements and 7,628 nodes (b).}\label{fig:cpu}
\end{figure*}

The performance of the proposed ensemble solver was first tested on multiple CPUs. 
We use distributed sparse matrices in the Yale compressed sparse row (CSR) format~\cite{bulucc2009parallel} with dense coefficient entries of size $9\cdot n_{s}$, where $n_{s}$ is the number of material property realizations and a local element matrix of size 3$\times$3 results from selecting a three-d.o.f.s shell finite element.
The code for the finite element assembly and sparse matrix-vector multiplication was developed in Cython+MPI+openMP and compared both with a C implementation and with the \textit{mkl\_cspblas\_dcsrgemv} routine provided through the Intel MKL library, using single and multiple threads. We verified the satisfactory performance of our implementation under a wide range of mesh sizes, number of cores, and with/without multithreading. Encouraging speedups were obtained on multiple CPUs, as shown in Figure~\ref{fig:matProd} and Figure~\ref{fig:cpu}.

\begin{figure*}[!ht]
\centering
\begin{subfigure}[b]{0.32\linewidth}
\centering
\includegraphics[width=\linewidth]{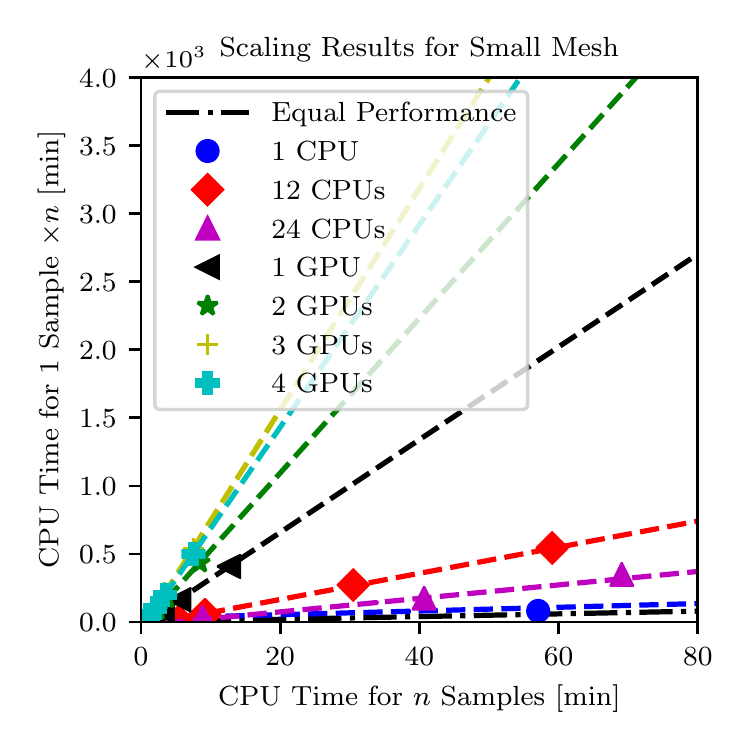}
\caption{}
\end{subfigure}
\begin{subfigure}[b]{0.32\linewidth}
\centering
\includegraphics[width=\linewidth]{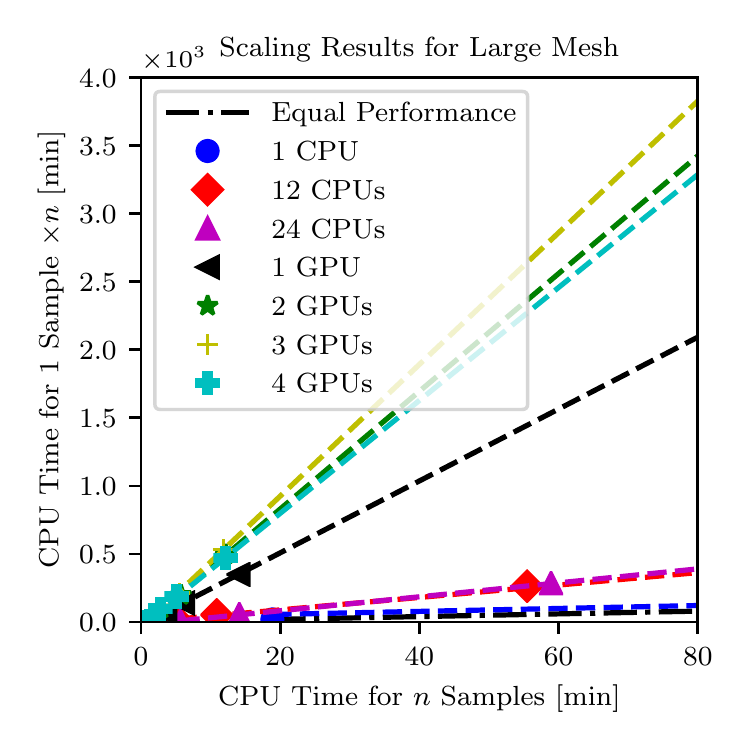}
\caption{}
\end{subfigure}
\caption{Performance of explicit ensemble solver on multiple CPUs/GPUs for a mesh with 5,074 elements, 2,565 nodes (a) and for a mesh with 15,136 elements and 7,628 nodes (b).}\label{fig:gpu}
\end{figure*}

\subsection{Distributed explicit FSI solver on multiple GPUs}\label{sec:fsiGPU}

\noindent We developed an hardware-independent openCL implementation of the solver running on multiple GPUs.
We started with a na\"ive CSR-based parallel sparse matrix-vector product (SpMV), known as \emph{CSR-Scalar}, where each row of the sparse matrix is assigned to a separate thread~\cite{garland2008sparse}. 
This works well on CPUs, but causes uncoalesced, slow memory accesses on GPUs, since elements in each row occupy consecutive addresses in memory, but consecutive threads access elements on different rows. In addition, long rows lead to an unequal amount of work among the threads and some of them need to wait for others to finish.
We then transitioned to a \emph{CSR-Vector} scheme~\cite{bell2009implementing}, assigning a wavefront (or so-called \emph{warp} on NVIDIA architectures) to work on a single row of the matrix. This allows for access to consecutive memory locations in parallel, resulting in fast coalesced loads. However, CSR-Vector can lead to poor GPU occupancy for short rows due to unused execution resources.
Improved performance can be achieved using \emph{ CSR-Stream}~\cite{greathouse2014efficient} which statically fixes the number of nonzeros that will be processed by one wavefront and streams all of these values into the local scratchpad memory, effectively utilizing the GPU’s DRAM bandwidth and improving over CSR-Scalar. 
CSR-Stream also dynamically determines the number of rows on which each wavefront will operate, thus improving over CSR-Vector.
While CSR-Stream substantially improves the performance of the spMV product kernel, the CPU-to-GPU data transfer still dominates the time step update, as shown in Figure~\ref{fig:gpuDataTransfer}.
This problem can be mitigated by sending data to the GPU in smaller chunks, thus overlapping data transfer and kernel execution. 

In addition, data transfer can be minimized by an assembly-free approach. 
Even though this is a standard practice in explicit finite element codes, it is particularly effective on a GPU for three reasons. First, storage of a sparse global matrix would occupy a significant portion of the GPU memory, posing restrictions on the model size. Second, indexing operations to access entries in the global matrix would cause severe uncoalesced memory access, reducing significantly the degree of parallelism in the GPU. Third, for the most common sparse matrix storage schemes, indexing always include searching, which would be particularly slow on GPU.
We compute local matrices directly in the GPU and take their product with a partition-based displacement vector, where only the displacements of shared nodes need to be synchronized through the root CPU.

To compute element matrices, we buffer data to the GPU before the beginning of the time integration loop, in order to avoid any CPU to GPU data transfer due to element assembly.
Buffered quantities include mesh geometry and material properties, particularly the product between the Young's modulus and thickness at each Gauss point for all random field realizations. Note how the rest of the local stiffness matrix is constant for linear triangular elements. In addition, the left-hand-side lumped mass matrix resulting from the central difference scheme does not change throughout the time loop.
We also leverage a mesh coloring algorithm, allowing working units to process different elements at the same time.
During each time step, each computing group in the GPU works on one element, while the working units in the same group work on different realizations and therefore have access to coalesced GPU memory.
Communication is only triggered by displacement synchronization. We use pinned host memory to speed up the data transfer between CPU and GPU.
All displacements for each realization and the local matrices are stored in private memory since they do not need to be shared with other working units.
Final GPU speed ups are illustrated in Figure~\ref{fig:gpu}.

\subsection{A Python code-base}\label{sec:code}

\noindent The CVFES solver is developed in Python 3 with optimization in Cython~\cite{behnel2010cython} and element assembly and matrix product implementation on openCL~\cite{stone2010opencl} (though the Python pyOpenCL library~\cite{klockner2012pycuda}).
%
The code leverages the VTK library~\cite{vtk} to read the solid, fluid mesh and boundary conditions. In this context the solver is fully compatible with the input files generated by the SimVascular software platform~\cite{lan2018re} and can be easily integrated with the SimVascular modeling workflow.
%
Partitioning on multiple CPUs and GPUs is obtained for both solvers using parMETIS~\cite{METIS}.
The code used to generate the results discussed in Section~\ref{sec:results} is available through a public GitHub repository at \url{https://github.com/desResLab/CVFES}. 

\section{Results}\label{sec:results}

\subsection{Ideal cylindrical benchmark}\label{sec:cylTestCase}

\noindent The first benchmark represents an ideal cylindrical lumen subject to aortic flow. 
The cylinder has a diameter equal to 4 cm and length of 30 cm, while two Mat\'ern random fields for thickness and elastic modulus have been assigned as discussed in Section~\ref{sec:randomField}. 
Specifically, a mean $\mu=7.0\times 10^{6}$ Barye and a standard deviation $\sigma=7.0\times 10^{5}$ Barye have been assumed for the elastic modulus, whereas the thickness random field is characterized through a mean $\mu=0.4$ cm and a standard deviation $\sigma=0.04$ cm. Three values of the correlation length were considered, equal to 0.95 cm, 3.7 cm and 7.2 cm, respectively (see discussion in~\cite{tran2019uncertainty}).
%
A uniform pressure of 13 mmHg was added to the pressure computed by the VMS fluid solver. We considered diastole as a natural (unstressed) state and applied the difference between a diastolic pressure of 80 mmHg and the mean brachial pressure in a healthy subject (i.e., with systolic pressure equal to 120 mmHg).
This configuration is analyzed both under steady state and pulsatile flow conditions, under fully fixed structural boundary conditions at the two cylinder ends.

\subsubsection{Steady state analysis}\label{sec:cylTestCaseSteadyState}

\begin{figure}[!ht]
\centering
\begin{subfigure}[c]{0.48\linewidth}
\centering
\includegraphics[width=1.0\textwidth]{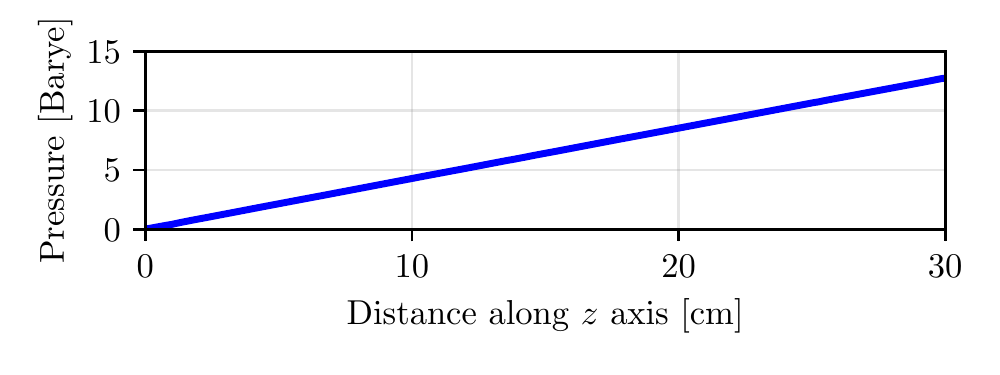}
\caption{}
\end{subfigure}
\begin{subfigure}[c]{0.48\linewidth}
\centering
\includegraphics[width=1.0\textwidth]{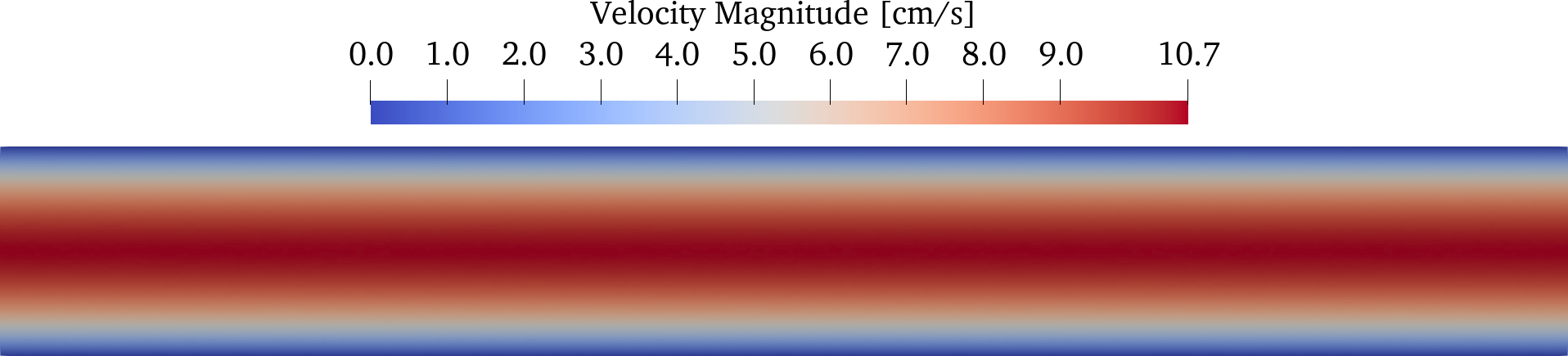}
\caption{}
\end{subfigure}
\caption{Steady state pressure (a) and velocity (b) distributions from variational multi-scale fluid solver.}\label{fig:cylSteadySol}
\end{figure}

\noindent The fluid solution is computed with the VMS fluid solver discussed in Section~\ref{sec:VMS} using a parabolic velocity profile at the inflow corresponding to a $-66.59$ mL/s volumetric flow rate, zero-traction boundary condition at the outlet and a no-slip condition at the lumen wall. 
As expected, the fluid solution show a perfectly linear relative pressure profile along the cylinder center path and a uniform parabolic velocity profile from inlet to outlet, typical of viscous-dominated Poiseuille flow, as shown in Figure~\ref{fig:cylSteadySol}.

The steady state pressure resulting from the fluid solver is applied as discussed in Section~\ref{sec:FSI} and 100 thickness and elastic modulus realizations are solved simultaneously.
The explicit structural simulation is run for $0.5$ seconds, until a steady state was observed.
Three wall mesh densities (coarse, medium and fine) are finally considered, consisting of 5,074, 15,136 and 32,994 triangular shell elements, with explicit time steps set to $\Delta t = 4.0\times 10^{-5}$ s, $\Delta t = 4.0\times 10^{-5}$ s and $\Delta t = 1.0\times 10^{-5}$ s, respectively and no viscous damping.

\begin{figure*}[!ht]
\centering
\includegraphics[width=0.32\textwidth]{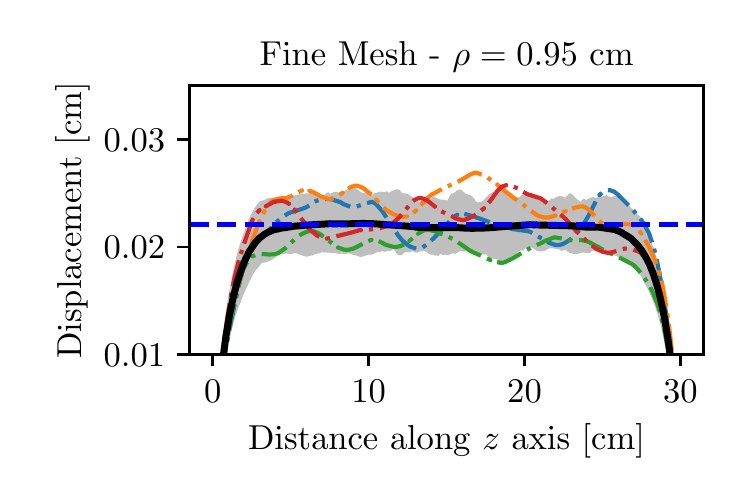}
\includegraphics[width=0.32\textwidth]{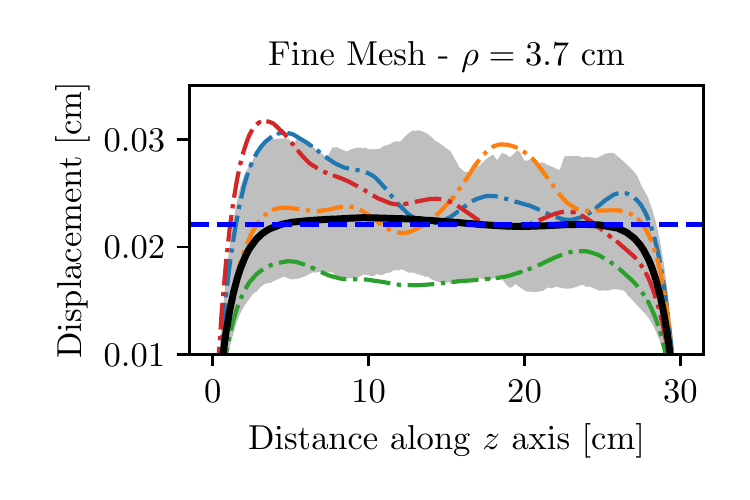}
\includegraphics[width=0.32\textwidth]{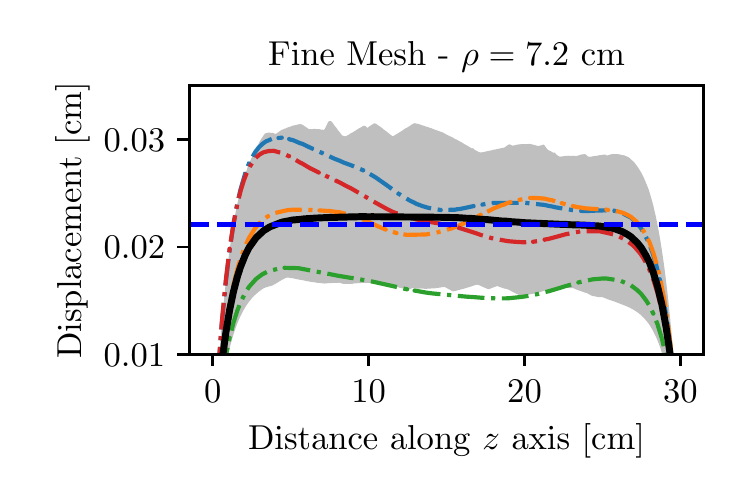}\\

\vspace{-10pt}

\includegraphics[width=0.32\textwidth]{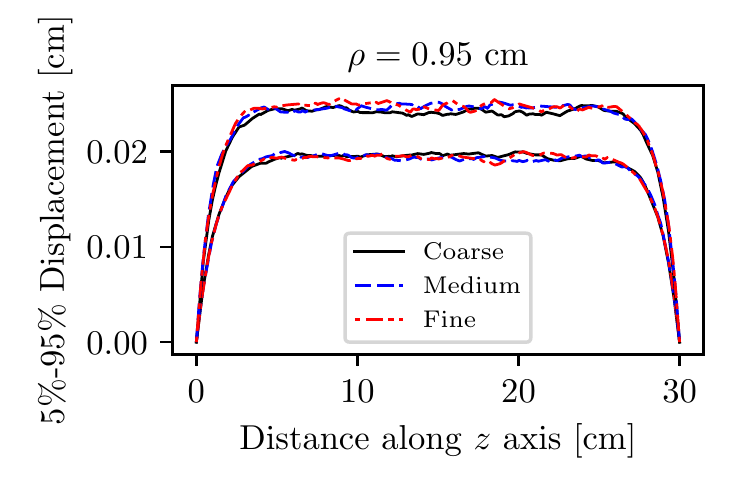}
\includegraphics[width=0.32\textwidth]{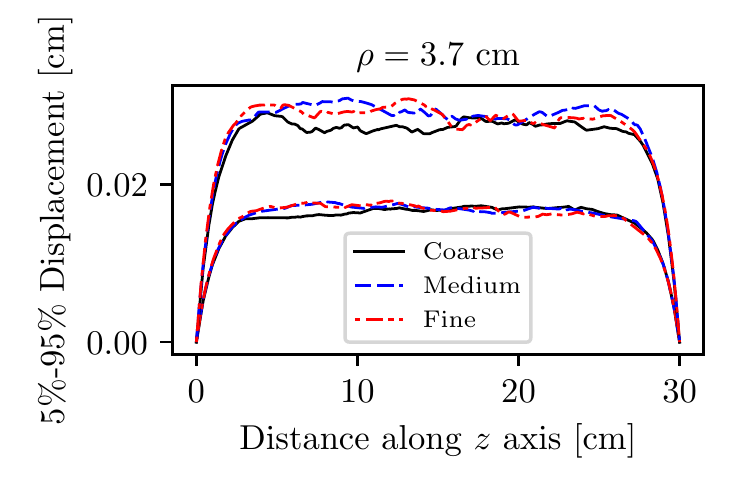}
\includegraphics[width=0.32\textwidth]{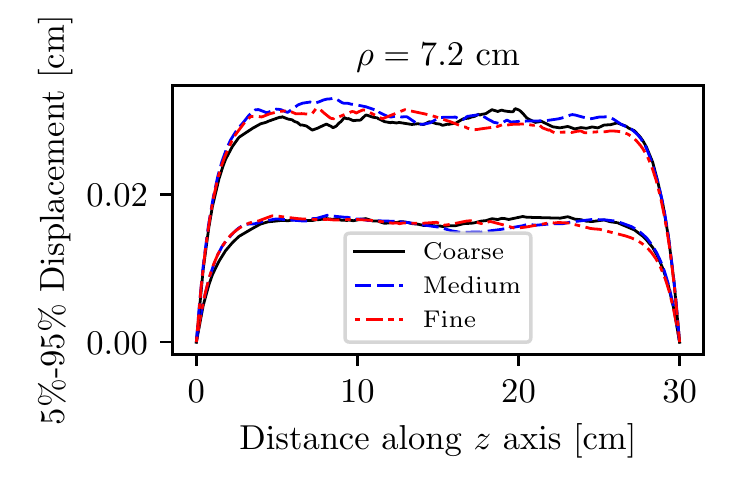}
\caption{Displacement magnitude along cylinder generator (longitudinal $z$ axis) for various correlation lengths (first row). 5\%-95\% confidence intervals for displacement magnitudes for three increasing mesh densities (i.e., coarse, medium and fine, second row).}\label{fig:cylSSDisp}
\end{figure*}

Displacement magnitudes along the longitudinal $z$ axis (cylinder generator) are shown in the top row of Figure~\ref{fig:cylSSDisp} for the finer mesh and various correlation lengths.
The mean displacement one diameter away from the fully fixed ends (thick black line) is consistent with an homogeneous solution for a thick cylinder with average elastic modulus and thickness (blue dashed line).
Displacements associated with single random field realizations are also shown, in Figure~\ref{fig:cylSSDisp} (top row), using colors.
As expected, the displacement wave length increases with the correlation length, and so does the displacement uncertainty quantified through the 5\%-95\% confidence interval (gray shaded area).
Finally, the second row of Figure~\ref{fig:cylSSDisp} shows how increasing the mesh density produces a limited difference in the 5\%-95\% confidence interval.

Circumferential stress was found to be the most significant component, as expected, showing uncertainty increasing with the correlation length, similarly to what observed for the displacement magnitude. 
The in-plane and out-of-plane shear components ($\sigma_{\theta z}$ and $\sigma_{rz}$), though much smaller than circumferential and axial stresses, are essentially related to the material property non-homogeneity and reduce for an increasing correlation length.
These stress components are zero both on average and by solving the model with average material properties. Thus, they can only be captured by explicitly modeling the spatial variability of material properties as in the proposed approach. 

\begin{figure*}[!ht]
\centering
\includegraphics[width=0.32\textwidth]{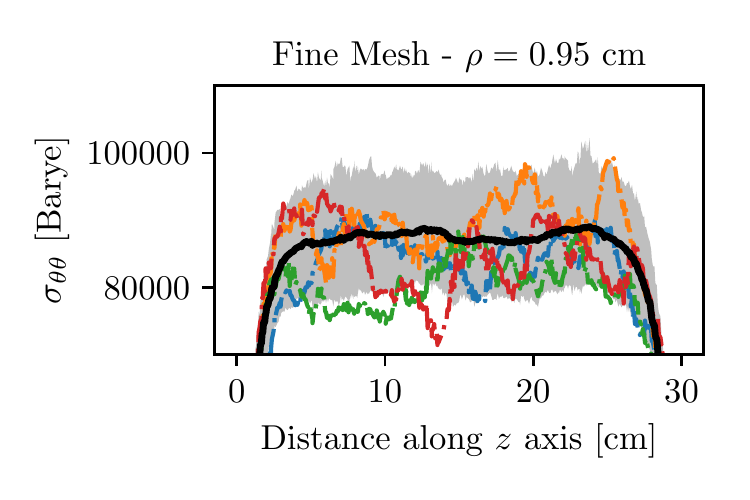}
\includegraphics[width=0.32\textwidth]{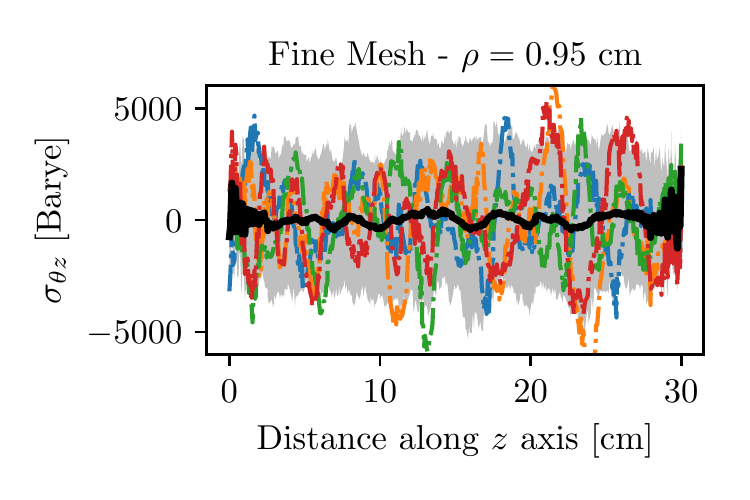}
\includegraphics[width=0.32\textwidth]{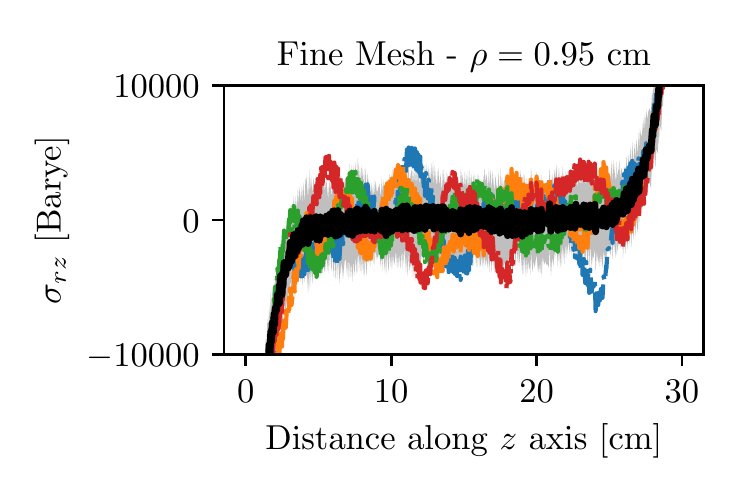}\\ 

\vspace{-10pt}

\includegraphics[width=0.32\textwidth]{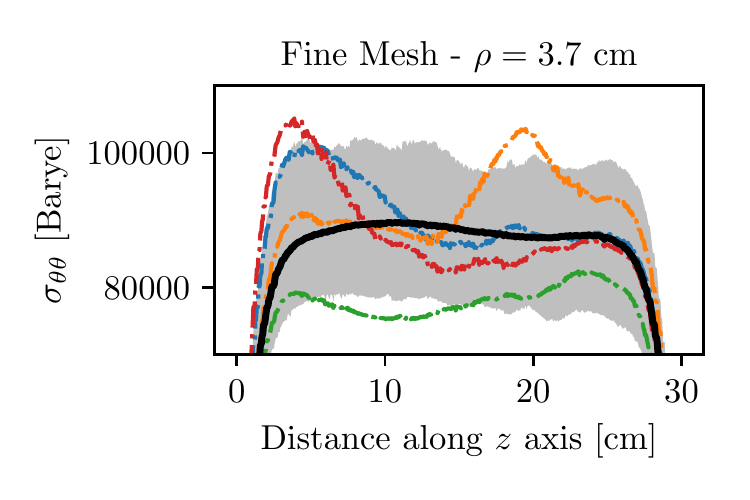}
\includegraphics[width=0.32\textwidth]{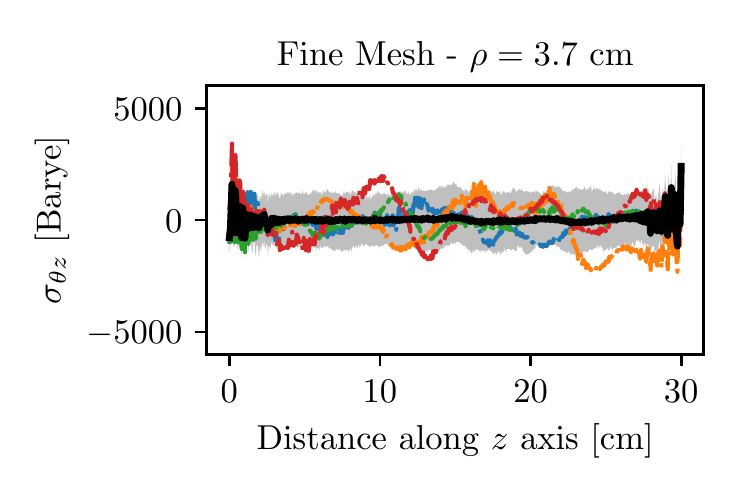}
\includegraphics[width=0.32\textwidth]{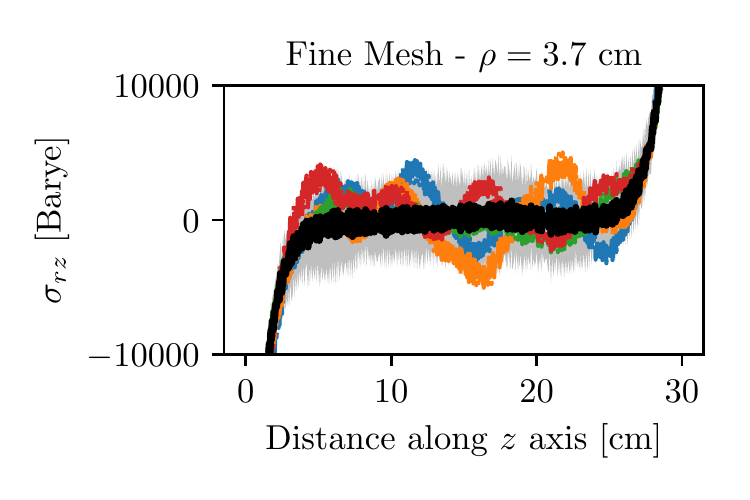}\\

\vspace{-10pt}

\includegraphics[width=0.32\textwidth]{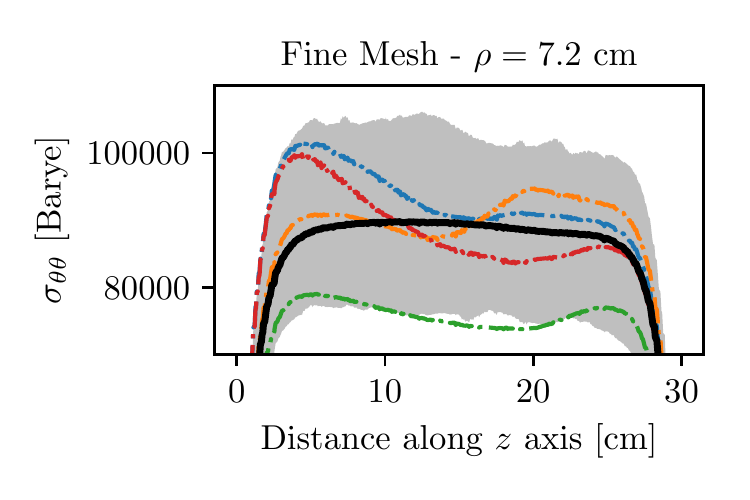}
\includegraphics[width=0.32\textwidth]{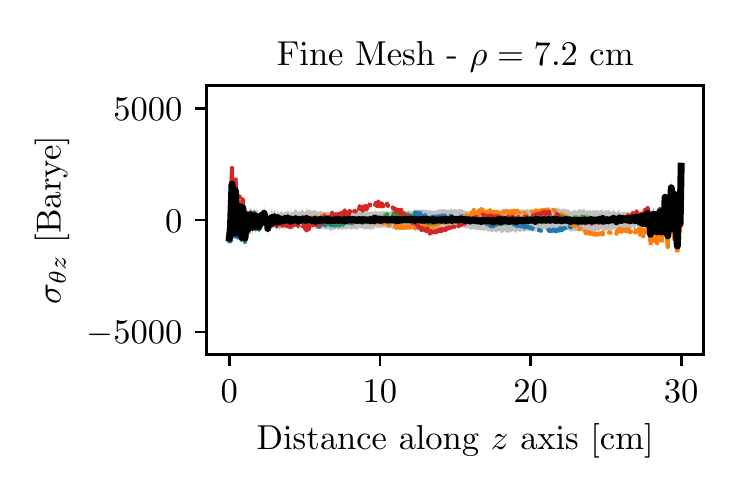}
\includegraphics[width=0.32\textwidth]{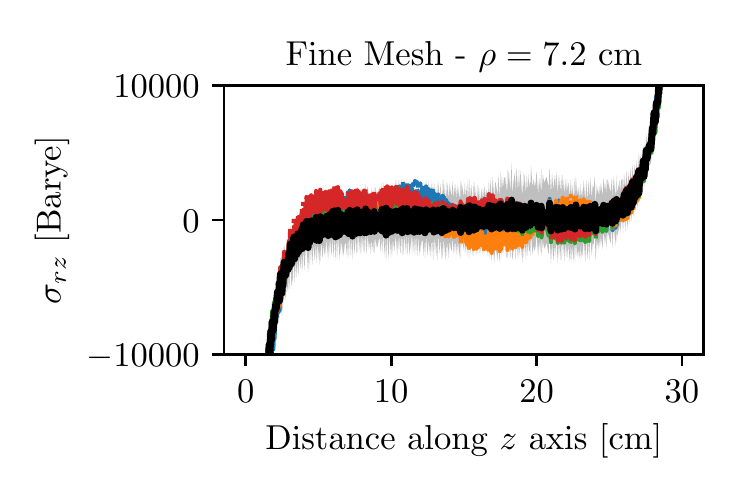}
\caption{Ensemble means, 5\%-95\% confidence intervals and single realizations of longitudinal stress profiles for various mesh densities and random field correlation lengths.}\label{fig:cylSSEnvStress}
\end{figure*}

\subsubsection{Validation under pulsatile flow conditions}\label{sec:cylTestCasePulsatile}

\begin{figure*}[!ht]
\centering
\begin{subfigure}[b]{0.32\linewidth}
\centering
\includegraphics[width=\textwidth]{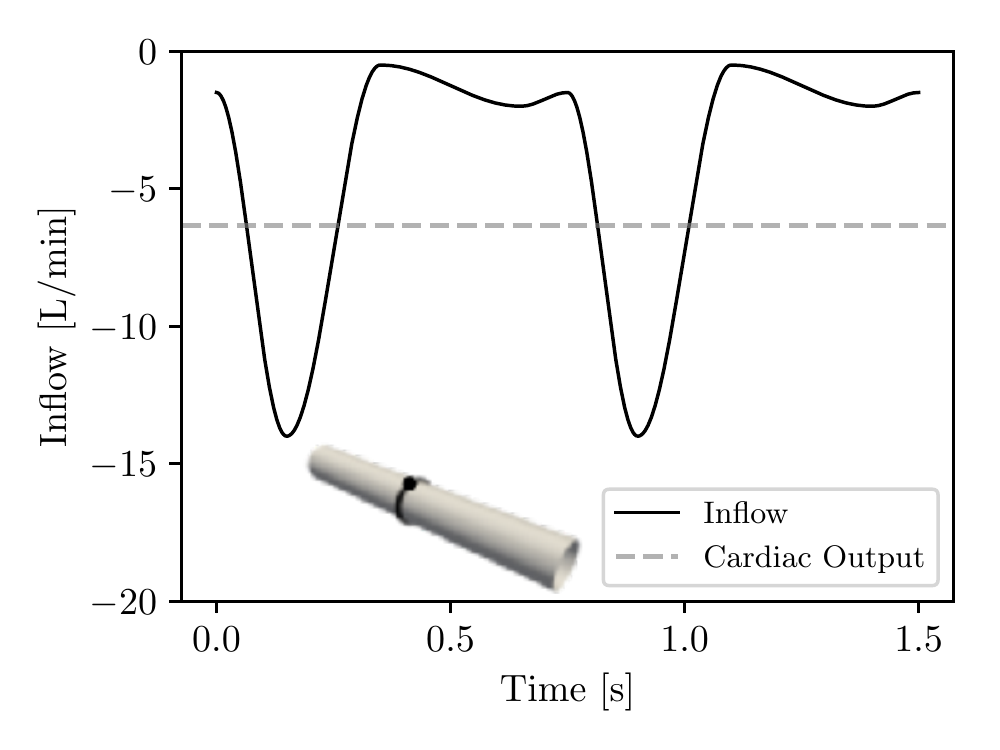}
\caption{}\label{fig:cylPuls_a}
\end{subfigure}
\begin{subfigure}[b]{0.32\linewidth}
\centering
\includegraphics[width=\textwidth]{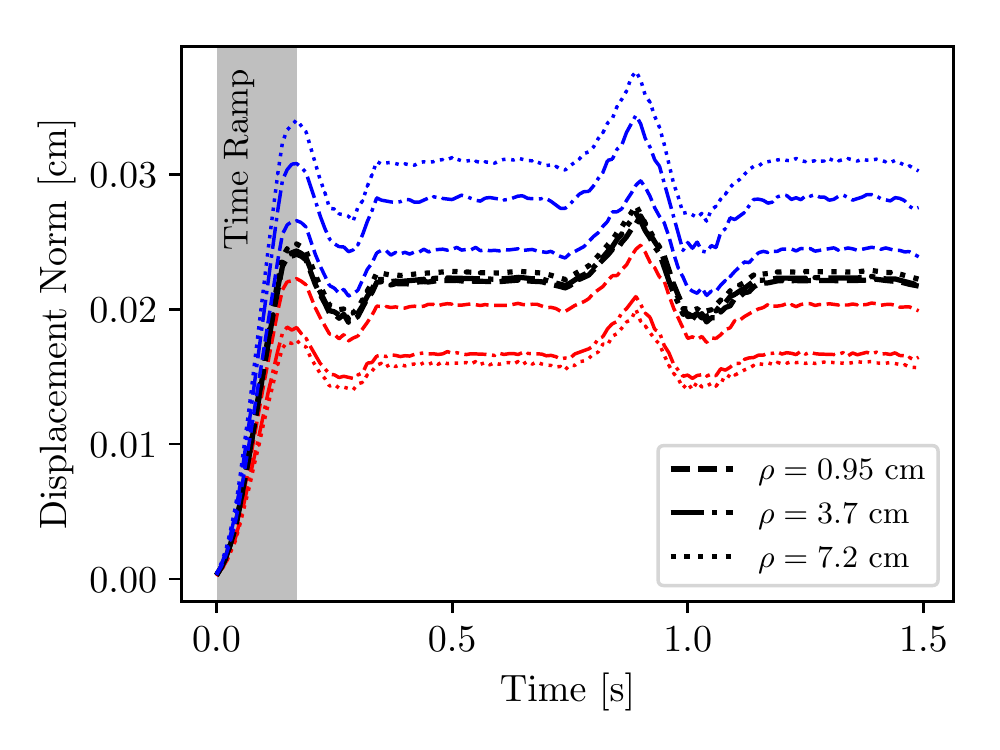}
\caption{}\label{fig:cylPuls_b}
\end{subfigure}
\begin{subfigure}[b]{0.32\linewidth}
\centering
\includegraphics[width=\textwidth]{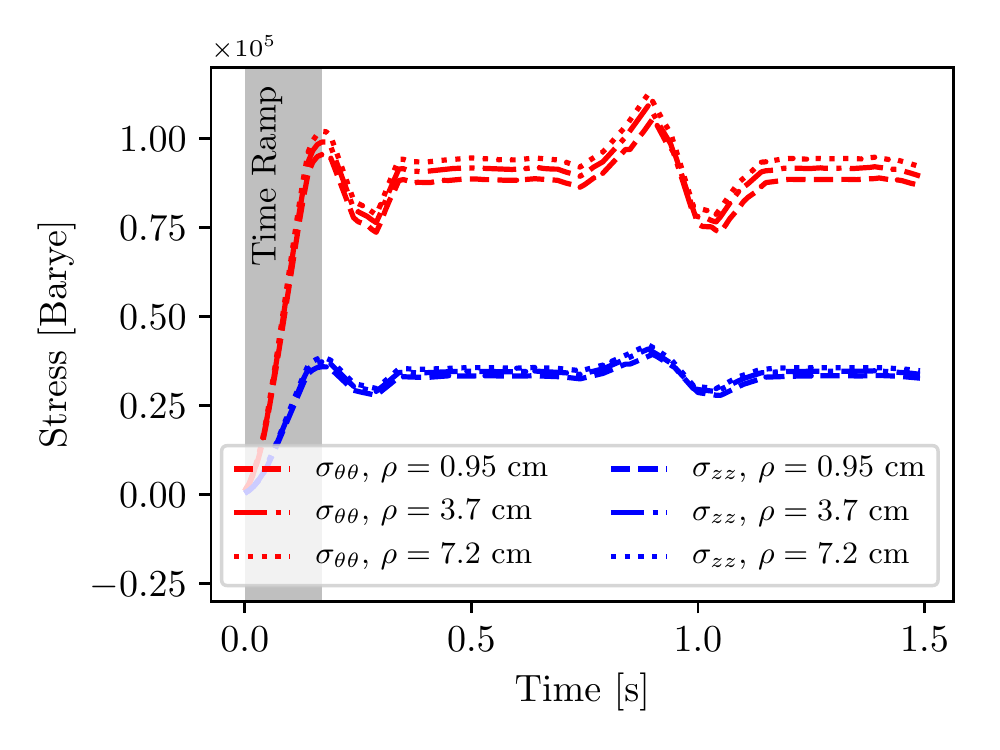}
\caption{}\label{fig:cylPuls_c}
\end{subfigure}
\caption{Results for pulsatile flow on ideal cylindrical vessel. Inflow time history and spatial location for acquisition of displacement and stress outputs (a). Displacement time history at selected spatial location with thick lines representing ensemble averages, and thin lines used for 5\%/95\% percentiles (b). Circumferential and axial stress time history at selected spatial location (c).}\label{fig:cylPuls}
\end{figure*}

\noindent A parabolic pulsatile inflow (Figure~\ref{fig:cylPuls_a}) was applied to the same cylindrical geometry discussed in the previous section, while all the other boundary conditions (walls and outlets) were kept the same as in the steady case.
The time step is set to $1.0\times 10^{-5}$ and the simulation run for two heart cycles and 100 material property realizations.
To avoid the application of impulsive loads which can excite a broad range of frequencies, significantly affecting the undamped dynamics, a ramp was applied to the wall loads resulting from the fluid solver in the last heart cycle. The ramp follows a sine wave and is kept active for the first $0.2$ seconds of the simulation. No damping was applied to the simulation, i.e., $\bm{f}_{v}=0$.

As expected, the resulting displacements follow a time profile similar to the inflow, and the 5\%-95\%confidence intervals increase with the correlation length of the underlying random field. 
Circumferential and axial stress components are the dominant stress components, which also increase with the correlation length, as observed for the steady state results. 

\subsection{Benchmark on patient-specific coronary model}\label{sec:coronaryTestCase}

\noindent We also demonstrate the results of the proposed ensemble solver on a patient-specific model of the left anterior descending coronary branch.
An ensemble of 100 model solutions were obtained using random field parameters for the elastic modulus equal to $\mu = 1.15\times 10^{7}$ Barye and $\sigma = 1.7\times 10^{6}$ Barye. For the thickness, we considered a mean equal to $\mu = 0.075$ cm and a standard deviation of $\sigma = 0.017$ cm.
A slip-free boundary condition was applied at the outlets of the fluid domain, whereas fully fixed mechanical restraints (i.e., all three nodal translations) where applied along the edges at both the inlets and outlets.
A uniform pressure has been superimposed to the lumen stress obtained from the fluid solver as discussed for the ideal aortic model in Section~\ref{sec:cylTestCase}.

\subsubsection{Steady state analysis}\label{sec:coronaryTestCaseSteadyState}

\begin{figure}[!ht]
\begin{subfigure}[c]{0.5\linewidth}
\centering
\includegraphics[width=\linewidth]{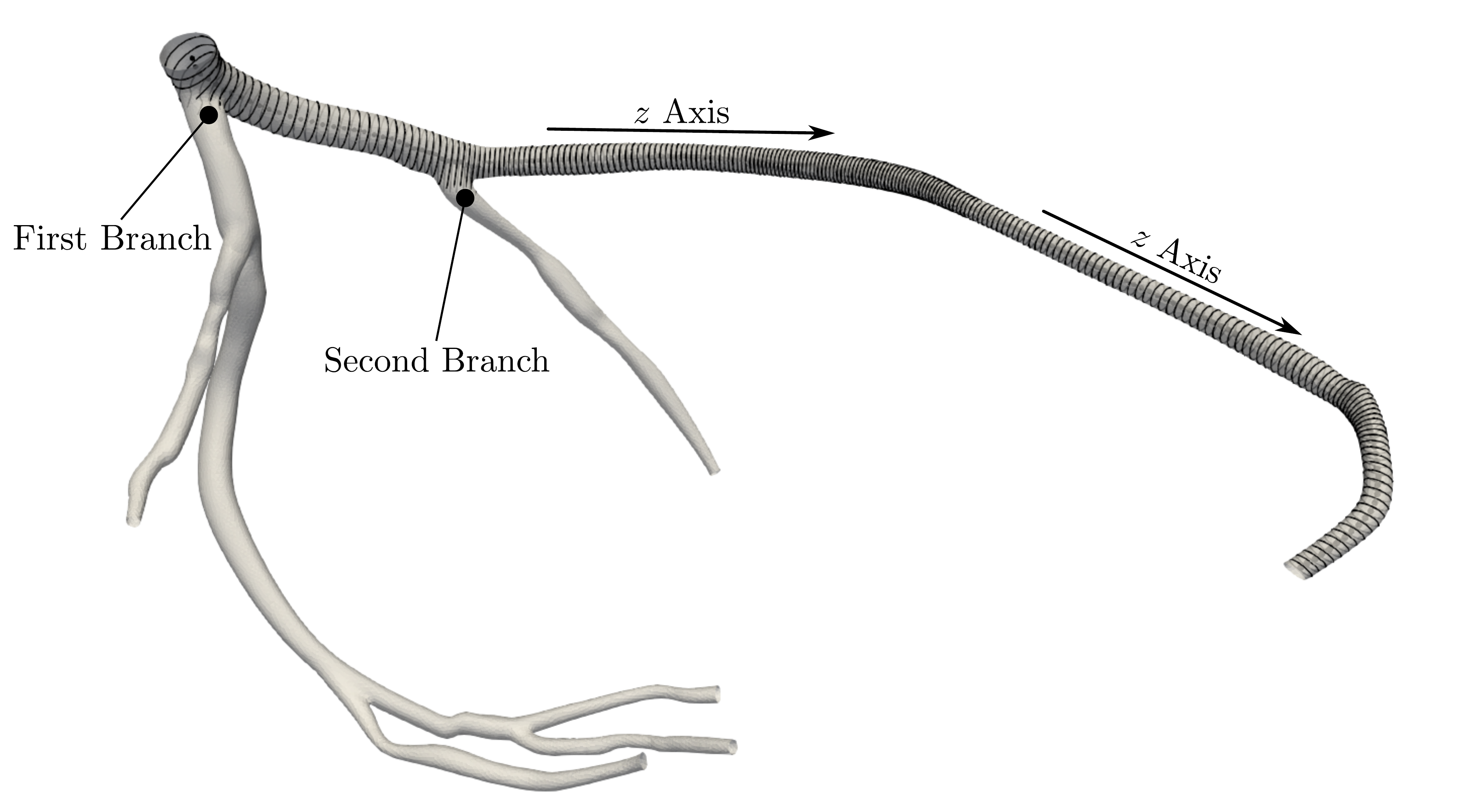}
\end{subfigure}
\begin{subfigure}[c]{0.5\linewidth}
\centering
\includegraphics[width=\linewidth]{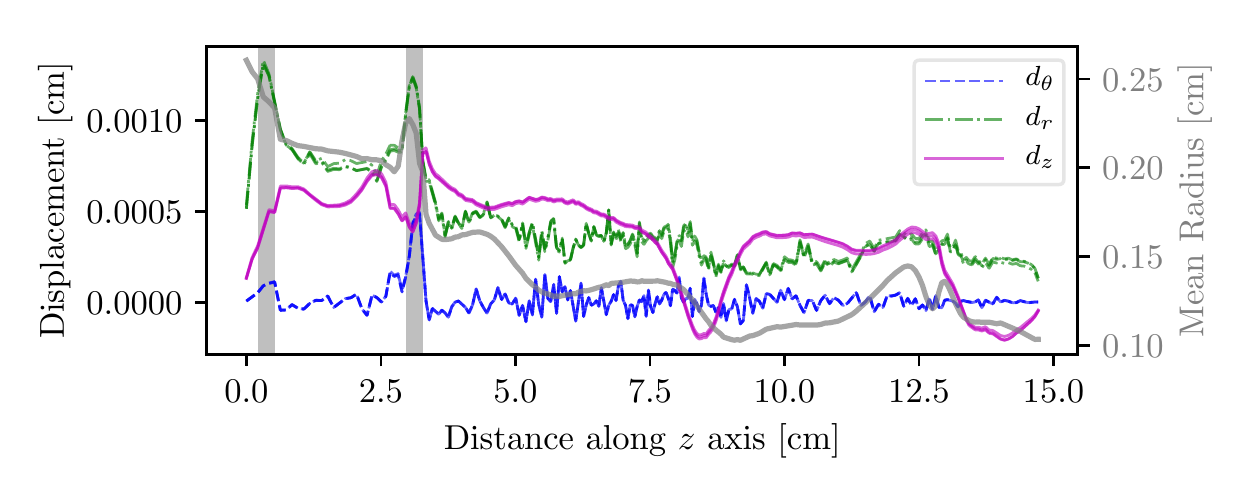}
\includegraphics[width=\linewidth]{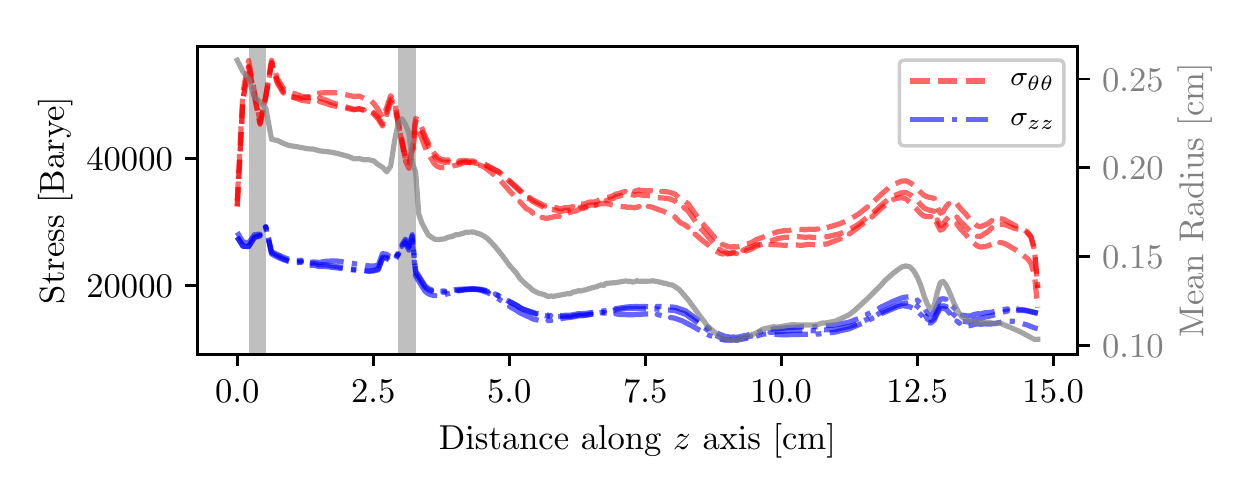}
\end{subfigure}
\caption{Displacement and stress profiles in left coronary artery LAD branch under steady flow, averaged over all material property realizations and cross-sectional slice.}\label{fig:lc_ss_res}
\end{figure}

\noindent A constant flow rate equal to $-0.28$ mL/s is applied at the model inlet with a parabolic profile, while a zero-traction boundary condition is applied at the outlets and a no-slip condition at the walls.
The explicit time step is set to $2.5\times 10^{-6}$ and the model was run for $0.13$ seconds to reach the steady state. No additional viscous force $\bm{f}_{v}$ was considered.
Figure~\ref{fig:lc_ss_res} shows the displacements and stress for all three analyzed correlation lengths, averaged through a cross sectional slice of the branch of interest, and plotted for successive slices along the longitudinal $z$ axis.

Displacement results confirm the absence of torsion, and a prevalent radial deformation mode, with rigid body motion evident from the axial displacements $d_{z}$, affected by the centerline path geometry. 
The stress results confirm the importance of the circumferential followed by the axial component. The circumferential stress reduces with the coronary branch radius as intuitively suggested by the Barlow (or Mariotte) formula for thin-walled cylinders.
Similar cylindrical displacements, circumferential and axial stress are observed for all three correlation lengths. For the selected coronary branch, the smaller correlation length (0.95 cm) is approximately equal to twice the largest diameter, inducing minimal changes in local deformability.

\subsubsection{Validation under pulsatile flow conditions}\label{sec:coronaryTestCasePulsatile}

\begin{figure}[!ht]
\begin{subfigure}[c]{0.5\linewidth}
\centering
\includegraphics[width=\linewidth]{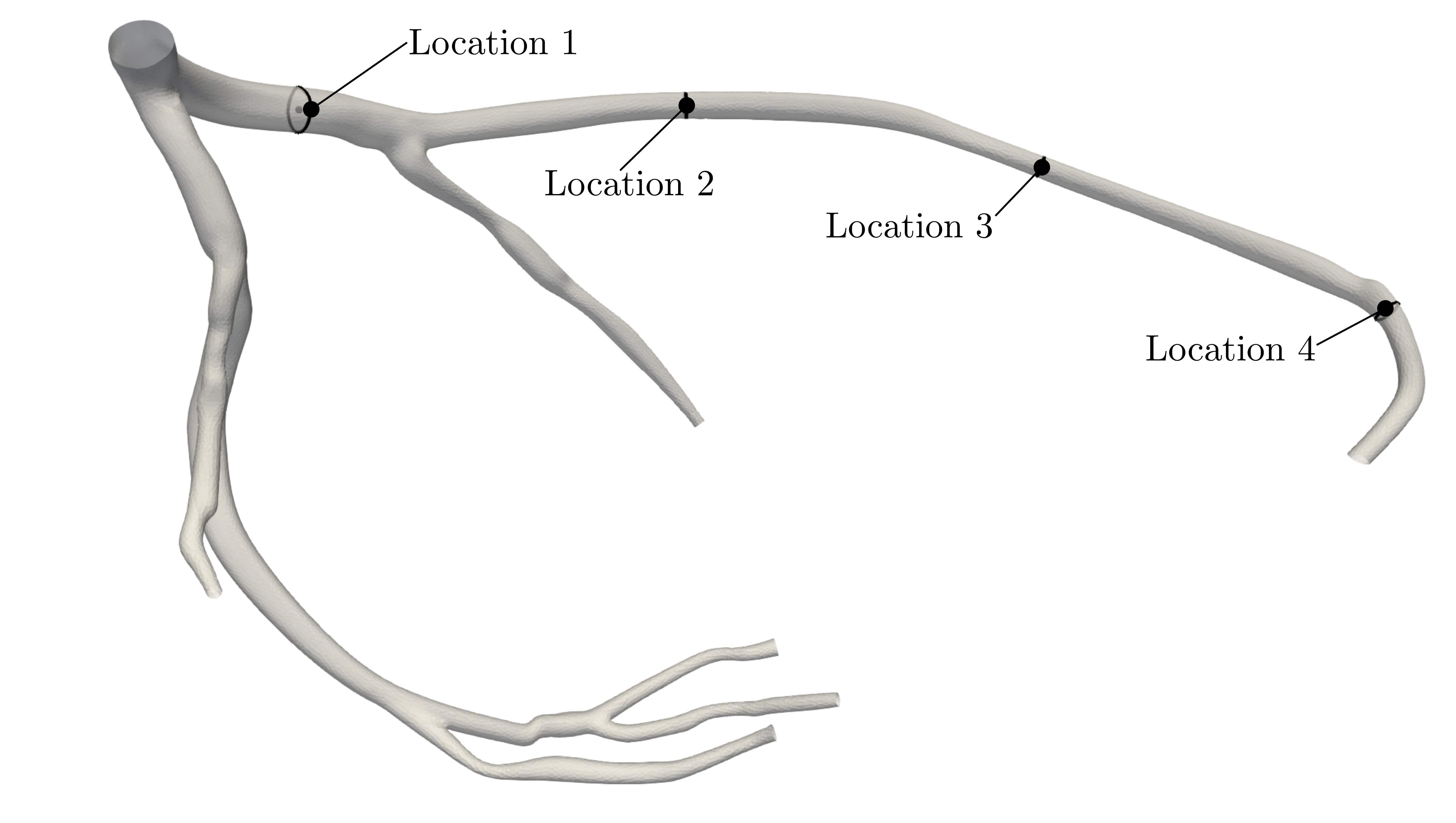}
\end{subfigure}
\begin{subfigure}[c]{0.5\linewidth}
\includegraphics[width=\linewidth]{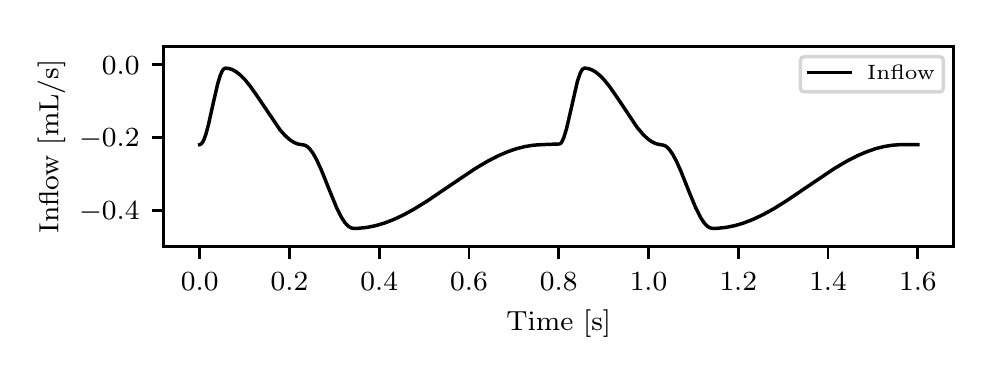}
\includegraphics[width=\linewidth]{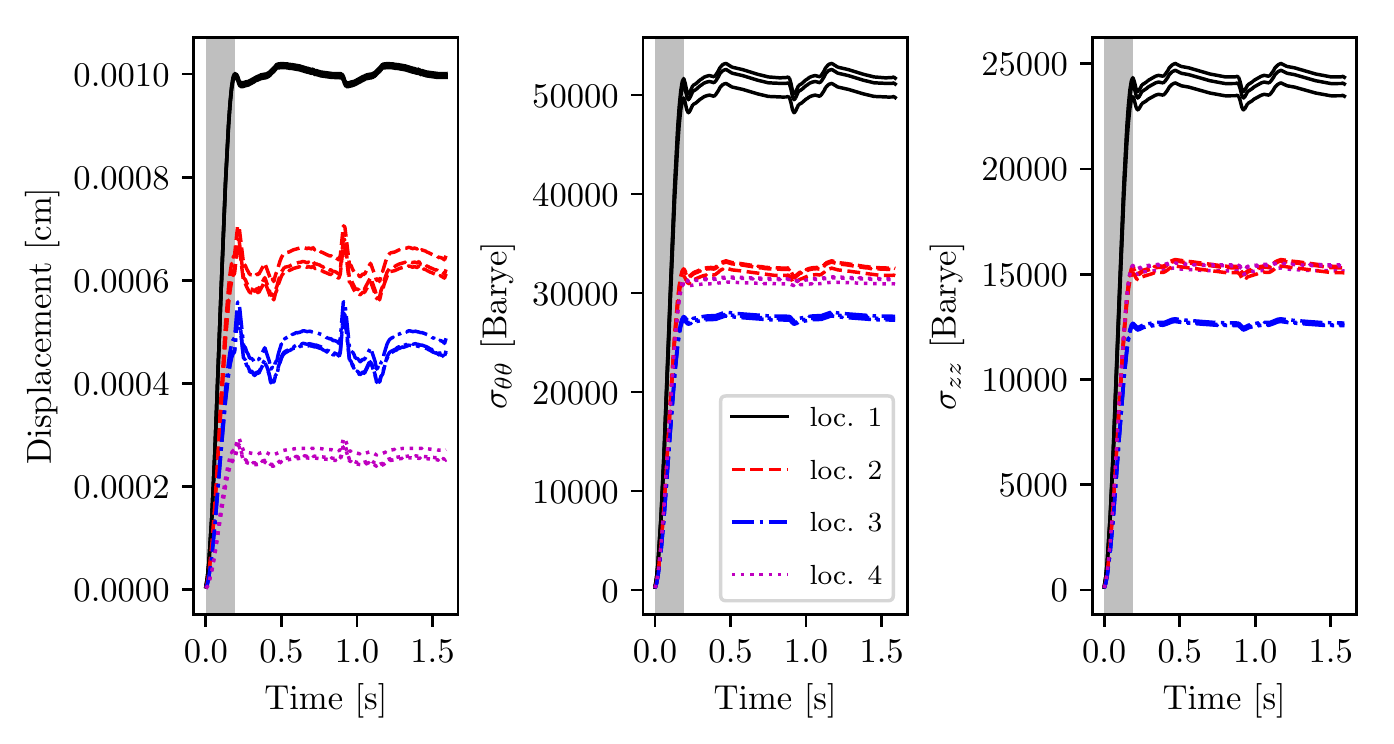}
\end{subfigure}
\caption{Displacement and stress profiles in left coronary artery LAD branch under pulsatile flow, averaged over all material property realizations and cross-sectional slice.}\label{fig:lcPulsResult}
\end{figure}

The same geometry analyzed in the previous section is subject to a parabolic pulsatile inflow shown in Figure~\ref{fig:lcPulsResult}.
A time step equal to $2.0\times 10^{-6}$ is selected, and the model is run for two complete heart cycles (1.6 seconds) and 100 material property realizations.  
Similar to the pulsatile cylindrical test case, a time ramp is applied during the first 0.2 s, to avoid impulsive loads produced by a non-zero wall stress at $t=0$, and a pressure of 13 mmHg is superimposed to the fluid wall stress.
A viscous force $\bm{f}_{v}$ was applied, using a damping coefficient equal to $c_{d} = 0.005$, which was found from various tests to remove the high-frequency oscillations without affecting the system dynamics.
Results for the average displacements and stress are shown in Figure~\ref{fig:lcPulsResult} at four successive locations over the center path of the LAD branch. Even for this case the circumferential and axial stress are the most relevant components and their time history is similar to the inflow and exhibit a maximum at diastole.

\subsection{Performance assessment}\label{sec:solverPerformance}

\noindent In this section, we compare the performance obtained by running the proposed ensemble solver on three cylindrical models with an increasing number of elements, as shown in Table~\ref{tab:small}, Table~\ref{tab:medium} and Table~\ref{tab:large}.
The explicit time step was set to $1.0\times 10^{-5}$ s for all models, and each run consisted of 1,000 time steps.
The GPUs used for these tests are four \emph{GeForce RTX 2080 Ti} with 11GB of RAM equipped with 4352 NVIDIA CUDA Cores and connected to the server main board through PCI express ports. 

Two types of speedup are investigated, the first relates to solving the same number of material property realizations on an increasing number of processors, either CPUs or GPUs, and provides an idea of the effectiveness of the various optimizations presented in Section~\ref{sec:fsiCPU} and Section~\ref{sec:fsiGPU}.
This speedup (first number in parenthesis) is observed to increase with the model size and the number of random field realizations.
The computation/communication tradeoff and the small mesh sizes selected for these tests also justify the negligible benefits of using 24 CPU cores instead of 12.
%
The speedup achieved by our GPU implementation is instead very relevant, i.e., approximately three orders of magnitude with respect to a single CPU implementation.

The second type of speedup quantifies the efficiency in the proposed ensemble solver, i.e., how much faster one can obtain the solution of multiple realizations by solving them at the same time, with respect to the solution of a single realization on the same hardware, multiplied by the total number of realizations.
The computational savings of an ensemble solution are quantified between one and two orders of magnitude, confirming our initial claim.

\begin{table*}[t]
\caption{Speedup - Model with 5074 elements, 2565 nodes - 1000 time steps with $\Delta t=1.0\times 10^{-5}$}\label{tab:small}
\resizebox{\linewidth}{!}{%
\begin{tabular}{llllllll}
\toprule
{\bf Hardware} & {\bf 1 Smp (Spd)} & {\bf 10 Smp (Spd)}  & {\bf 50 Smp (Spd)}  & {\bf 100 Smp (Spd)} & {\bf 200 Smp (Spd)} & {\bf 500 Smp (Spd)} & \\
\midrule
{\bf 1 CPU} & 0:00:16 (1.0,1.0) & 0:01:54 (1.43,1.0) & 0:09:46 (1.39,1.0) & 0:18:53 (1.44,1.0) & 0:37:20 (1.46,1.0) & 1:40:13 (1.36,1.0)\\
{\bf 12 CPU} & 0:00:10 (1.0,1.50) & 0:00:18 (5.92,6.20) & 0:01:00 (8.92,9.62) & 0:01:58 (9.21,9.59) & 0:03:45 (9.65,9.94) & 0:09:35 (9.46,10.45)\\
{\bf 24 CPU} & 0:00:06 (1.0,2.35) & 0:00:17 (3.93,6.45) & 0:01:21 (4.27,7.22) & 0:02:18 (5.03,8.21) & 0:05:25 (4.26,6.87) & 0:11:38 (4.98,8.61)\\
\midrule
{\bf 1 GPU} & 0:00:01 (1.0,9.42) & 0:00:01 (10.20,67.19) & 0:00:02 (29.82,201.96) & 0:00:05 (32.75,214.30) & 0:00:09 (37.84,244.67) & 0:00:20 (42.31,293.77)\\
{\bf 2 GPU} & 0:00:01 (1.0,10.56) & 0:00:01 (9.17,67.69) & 0:00:02 (36.27,275.28) & 0:00:03 (44.01,322.76) & 0:00:05 (55.86,404.83) & 0:00:11 (66.74,519.27)\\
{\bf 3 GPU} & 0:00:01 (1.0,9.49) & 0:00:01 (9.53,63.25) & 0:00:01 (44.86,305.92) & 0:00:02 (59.66,393.12) & 0:00:04 (75.29,490.21) & 0:00:09 (91.75,641.43)\\
{\bf 4 GPU} & 0:00:01 (1.0,9.34) & 0:00:01 (9.70,63.35) & 0:00:01 (46.12,309.52) & 0:00:02 (58.67,380.35) & 0:00:04 (75.09,481.08) & 0:00:09 (90.21,620.49)\\
\bottomrule
\end{tabular}}
\vspace{3pt}
\footnotesize{(*) All time entries are in a \emph{days:hours:minutes:seconds} format. The speedup is indicated as $(x,y)$, where $x$ is the speedup with

\vspace{-3pt}

$\quad\,\,$ respect to the number of samples and $y$ the speedup by distributing the computation on multiple CPUs or GPUs.}
\end{table*}

\begin{table*}[t]
\caption{Speedup - Model with 15136 elements, 7628 nodes - 1000 time steps with $\Delta t=1.0\times 10^{-5}$}\label{tab:medium}
\resizebox{\linewidth}{!}{%
\begin{tabular}{llllllll}
\toprule
{\bf Hardware} & {\bf 1 Smp (Spd)} & {\bf 10 Smp (Spd)}  & {\bf 50 Smp (Spd)}  & {\bf 100 Smp (Spd)} & {\bf 200 Smp (Spd)} & {\bf 500 Smp (Spd)} & \\
\midrule
{\bf 1 CPU} & 0:00:37 (1.0,1.0) & 0:05:24 (1.16,1.0) & 0:25:33 (1.23,1.0) & 0:52:29 (1.20,1.0) & 2:03:22 (1.02,1.0) & 4:48:25 (1.09,1.0)\\
{\bf 12 CPU} & 0:00:10 (1.0,3.59) & 0:00:21 (4.83,14.89) & 0:01:50 (4.74,13.82) & 0:04:05 (4.29,12.83) & 0:07:23 (4.74,16.69) & 0:19:19 (4.54,14.92)\\
{\bf 24 CPU} & 0:00:11 (1.0,3.32) & 0:00:28 (4.02,11.48) & 0:01:57 (4.83,13.01) & 0:03:58 (4.77,13.21) & 0:07:20 (5.16,16.79) & 0:18:38 (5.08,15.47)\\
\midrule
{\bf 1 GPU} & 0:00:03 (1.0,9.72) & 0:00:03 (10.16,84.84) & 0:00:07 (26.56,209.58) & 0:00:13 (27.89,225.99) & 0:00:24 (31.22,297.26) & 0:00:56 (34.28,305.28)\\
{\bf 1 GPU} & 0:00:02 (1.0,15.01) & 0:00:02 (10.11,130.45) & 0:00:04 (29.17,355.60) & 0:00:08 (30.94,387.40) & 0:00:13 (37.20,547.27) & 0:00:30 (40.80,561.29)\\
{\bf 1 GPU} & 0:00:02 (1.0,17.02) & 0:00:02 (9.47,138.46) & 0:00:03 (29.37,405.87) & 0:00:06 (33.57,476.46) & 0:00:10 (40.38,673.50) & 0:00:24 (44.55,694.70)\\
{\bf 1 GPU} & 0:00:01 (1.0,20.01) & 0:00:02 (8.21,141.15) & 0:00:03 (28.94,470.08) & 0:00:05 (35.15,586.47) & 0:00:08 (42.02,823.93) & 0:00:19 (48.60,891.06)\\
\bottomrule
\end{tabular}}
\vspace{3pt}
\footnotesize{(*) All time entries are in a \emph{days:hours:minutes:seconds} format. The speedup is indicated as $(x,y)$, where $x$ is the speedup with 

\vspace{-3pt}

$\quad\,\,$ respect to the number of samples and $y$ the speedup by distributing the computation on multiple CPUs or GPUs.}
\end{table*}

\begin{table*}[t]
\caption{Speedup - Model with 131552 elements, 65896 nodes - 1000 time steps with $\Delta t=1.0\times 10^{-5}$}\label{tab:large}
\resizebox{\linewidth}{!}{%
\begin{tabular}{llllllll}
\toprule
{\bf Hardware} & {\bf 1 Smp (Spd)} & {\bf 10 Smp (Spd)}  & {\bf 50 Smp (Spd)}  & {\bf 100 Smp (Spd)} & {\bf 200 Smp (Spd)} & {\bf 500 Smp (Spd)} & \\
\midrule
{\bf 1 CPU} & 0:06:47 (1.0,1.0) & 0:56:58 (1.19,1.0) & 4:06:55 (1.38,1.0) & 7:55:06 (1.43,1.0) & - & -\\
{\bf 12 CPU} & 0:00:22 (1.0,18.13) & 0:02:03 (1.83,27.74) & 0:10:59 (1.71,22.48) & 0:21:32 (1.74,22.06) & 0:50:50 (1.46,1.0) & -\\
{\bf 24 CPU} & 0:00:23 (1.0,17.50) & 0:02:57 (1.32,19.29) & 0:13:34 (1.43,18.20) & 0:26:51 (1.45,17.69) & 0:54:27 (1.43,0.93) & 2:47:47 (1.16,1.0)\\
\midrule
{\bf 1 GPU} & 0:00:29 (1.0,13.93) & 0:00:30 (9.74,113.79) & 0:01:00 (24.17,244.68) & 0:02:00 (24.33,236.96) & 0:03:33 (27.48,14.32) & -\\
{\bf 2 GPU} & 0:00:16 (1.0,25.45) & 0:00:16 (9.95,212.32) & 0:00:31 (25.24,466.69) & 0:01:02 (25.54,454.35) & 0:01:53 (28.31,26.95) & 0:04:18 (30.99,38.94)\\
{\bf 3 GPU} & 0:00:11 (1.0,37.06) & 0:00:11 (9.68,300.82) & 0:00:21 (25.24,679.48) & 0:00:42 (25.69,665.52) & 0:01:17 (28.57,39.60) & 0:02:54 (31.48,57.60)\\
{\bf 4 GPU} & 0:00:08 (1.0,46.39) & 0:00:09 (9.52,370.13) & 0:00:17 (25.33,853.57) & 0:00:33 (26.47,858.07) & 0:00:59 (29.47,51.13) & 0:02:16 (32.24,73.82)\\
\bottomrule
\end{tabular}}
\vspace{3pt}
\footnotesize{(*) All time entries are in a \emph{days:hours:minutes:seconds} format. The speedup is indicated as $(x,y)$, where $x$ is the speedup with 

\vspace{-3pt}

\noindent$\quad\,\,$ respect to the number of samples and $y$ the speedup by distributing the computation on multiple CPUs or GPUs.}
\end{table*}

\section{Conclusions and future work}\label{sec:discussion}

\noindent Our study investigates the efficiency achievable by a ensemble cardiovascular solver on modern GPU architectures. The basic methodological approaches are discussed together with details on our efforts to optimize execution on such architectures, both algorithmically and in terms of host-to-device communication.
In particular, computation of local element matrices is performed directly on the GPU, and working units are used to determine displacement increments for independent material property realizations.

The main result from our analysis is the observation that explicit-in-time ensemble solvers based on matrix-vector product naturally achieve high scalability, due to their ability to generate large amount of computations and re-usable data patterns provided by independent realizations. 
This also suggests how ensemble cardiovascular solvers are \emph{ideal} to efficiently generate campaigns of high-fidelity model solutions that form a pre-requisite for uncertainty quantification studies, the state-of-the-art paradigm for the analysis of cardiovascular systems, due to their ability to quantify the effects of multiple sources of uncertainty on the simulation outputs.

We have also validated our solver for the steady state and pulsatile solution of an idealized aortic flow and a patient-specific model of the left coronary artery, under vessel wall mechanical property uncertainty, modeled through a Gaussian random field approximation.

Future work will be devoted to further improve the computational efficiency of the proposed approach and to transition from a segregated to a fully coupled Arbitrary Lagrangian-Eulerian fluid-structure interaction paradigm. In addition, the proposed approach only considered uncertainties due to vessel wall thickness and elastic modulus. We plan to support uncertainty also in the boundary conditions and model geometry.

\section*{Acknowledgements}
This work was supported by a National Science Foundation award \#1942662 \emph{CAREER: Bayesian Inference Networks for Model Ensembles} (PI Daniele~E.~Schiavazzi). This research used computational resources provided through the Center for Research Computing at the University of Notre Dame. We also acknowledge support from the open source SimVascular project at \url{www.simvascular.org}.

\section*{Conflict of interest}

\noindent The authors declare that they have no conflict of interest.

\bibliographystyle{unsrt}  
\bibliography{refs}

\end{document}